\documentclass{stacs_proc}
\usepackage{enumerate}
\usepackage{xspace}
\usepackage[latin1]{inputenc}
\usepackage[loose]{subfigure}
\usepackage{ifthen}
\usepackage{xcolor}
\usepackage{pstricks}
\input{dvipsnam.def}
\usepackage{pst-node}
\usepackage{pst-plot}
\usepackage{pst-coil}
\usepackage{multido}
\usepackage{pst-3d}
\usepackage{calc}

\newcommand{\SQRTwo}{0.717}

\newcommand{\IOL}[2]{#1\!\mid\! #2}

\newcommand{\SmallScale}{0.5}
\newcommand{\TinyScale}{0.42}
\newlength{\MediumStateDiameter}
\newlength{\SmallStateDiameter}
\newlength{\LargeStateDiameter}
\newlength{\VerySmallStateDiameter}
\newlength{\StateLineWidth}        
\newcommand{\StateLineStyle}{solid} 
\newcommand{\StateLineColor}{black}
\newif\ifStateLineDbl \StateLineDblfalse 
\newcommand{\StateLineDblCoef}{0.6} 
\newcommand{\StateLineDblSep}{0.4} 
\newcommand{\VSStateLineCoef}{.6} 
\newcommand{\StateFillStatus}{solid} 
\newcommand{\StateFillColor}{white}
\newcommand{\StateLabelColor}{black}
\newcommand{\StateLabelScale}{1.7}
\newcommand{\DimStateLineStyle}{solid} 
\newcommand{\DimStateLineCoef}{1} %
\newcommand{\DimStateLineColor}{gray}
\newcommand{\DimStateLabelColor}{gray}
\newcommand{\DimStateFillColor}{white}
\newlength{\EdgeLineWidth}
\newcommand{\EdgeLineStyle}{solid}
\newif\ifEdgeLineDbl \EdgeLineDblfalse 
\newcommand{\EdgeLineDblCoef}{0.5} 
\newcommand{\EdgeLineDblSep}{0.6} 
\newcommand{\EdgeLineColor}{black}
\newlength{\EdgeArrowWidth}
\newlength{\EdgeDblArrowWidth}
\newcommand{\EdgeArrowLengthCoef}{1.4}
\newcommand{\EdgeDblArrowLengthCoef}{1.7}
\newcommand{\EdgeArrowInset}{0.1}
\newcommand{\EdgeArrowStyle}{->}
\newcommand{\EdgeRevArrowStyle}{<-}
\newcommand{\EdgeLineBorderCoef}{2}
\newcommand{\EdgeLineBorderColor}{white}
\newcommand{\EdgeLabelColor}{black}
\newcommand{\EdgeLabelScale}{1.7}
\newcommand{\DimEdgeLineCoef}{1.2} 
\newcommand{\DimEdgeLineStyle}{solid} 
\newcommand{\DimEdgeLineColor}{gray}
\newcommand{\DimEdgeLabelColor}{gray}
\newlength{\ZZSize}
\newcommand{\ZZShape}{0.5}
\newcommand{\ZZLineWidth}{1.7}
\newcommand{\TransLabelZZCoef}{0.6}
\newlength{\EdgeOffset}
\newcommand{\VaucArcAngle}{15}
\newcommand{\VaucArcCurvature}{0.8}
\newlength{\VaucArcOffset}
\newcommand{\VaucLArcAngle}{30}
\newcommand{\VaucLArcCurvature}{0.8}
\newlength{\LoopOffset}
\newlength{\LoopVarOffset}
\newcommand{\LoopAngle}{30}
\newcommand{\CLoopAngle}{22}
\newcommand{\LoopVarAngle}{28}
\newcommand{\LoopOnMediumState}{7}
\newcommand{\LoopOnSmallState}{9.6} 
\newcommand{\LoopOnLargeState}{5.8}

\newcommand{\CLoopOnMediumState}{8}
\newcommand{\CLoopOnSmallState}{12}
\newcommand{\CLoopOnLargeState}{6}

\newlength{\TransLabelSep}
\newcommand{\EdgeLabelPosit}{.45}\newcommand{\EdgeLabelRevPosit}{.55}
\newcommand{\ArcLabelPosit}{.4}
\newcommand{\LArcLabelPosit}{.4}
\newcommand{\LoopLabelPosit}{.25}
\newcommand{\CLoopLabelPosit}{.25}
\newcommand{\InitStateLabelPosit}{.1}
\newcommand{\FinalStateLabelPosit}{.9}
\newcommand{\ArrowOnStateCoef}{}
\newcommand{\ArrowOnMediumState}{1.5}
\newcommand{\ArrowOnSmallState}{1.7} 
\newcommand{\ArrowOnLargeState}{1.3}
\newcommand{\ArrowOnVerySmallState}{5} 
\newlength{\VertShiftH} 
\newlength{\VertShiftD} 
\newlength{\VertShift}
\newif\ifVCFrame
\newcommand{\HideFrame}{\VCFramefalse}

\newif\ifVCGrid
\newcommand{\HideGrid}{\VCGridfalse}

\newif\ifVCRigidLabel
\newcommand{\RigidLabel}{\VCRigidLabeltrue}

\newif\ifVCStateLabelBaseLine

\newcommand{\FloatingLabel}{\VCStateLabelBaseLinefalse}
\HideFrame
\HideGrid
\RigidLabel
\FloatingLabel
\psset{unit=1cm}
\newpsstyle{VaucFrameStyle}{arrows=-,framesep=0pt,%
                     linewidth=0.6pt,linecolor=black,%
                                         linestyle=solid,%
                                         doubleline=false,%
                                         fillcolor=white,fillstyle=none,%
                                         cornersize=relative,framearc=0}
\newcommand{\FrameStyle}{\psset{style=VaucFrameStyle}}
\newpsstyle{VaucGridStyle}{%
      gridwidth=0.6pt,griddots=10,subgriddiv=1,%
          gridlabels=7pt}
\newcommand{\GridStyle}{\psset{style=VaucGridStyle}}
\newenvironment{VCPicture}[2][.5]%
  {\settoheight{\VertShiftH}{$\{$}%
   \settodepth{\VertShiftD}{$\{$}%
   \setlength{\VertShift}{.5\VertShiftD-.5\VertShiftH}%
   \begin{pspicture}[#1]#2%
   \ifVCFrame \FrameStyle \psframe#2\fi%
   \ifVCGrid \FrameStyle\GridStyle \psgrid#2\fi}%
  {\RstState\RstEdge%
   \end{pspicture}}
\newcommand{\VCScale}{0.6}
\newcommand{\FixVCScale}[1]{\renewcommand{\VCScale}{#1}}

\newcommand{\SmallPicture}{\FixVCScale{\SmallScale}}
\newcommand{\TinyPicture}{\FixVCScale{\TinyScale}}
\newif\ifVCName

\newcommand{\VCDraw}[2][\VCGridScale]{%
\psset{unit=#1cm}%
\scalebox{\VCScale}{#2}%
\psset{unit=1cm}}%
\newcommand{\VCPut}[3][0]{\rput{#1}#2{#3}}%
\newlength{\StateLineWid}
\newcommand{\StateLineSty}{\StateLineStyle} 
\newcommand{\StateLineCol}{\StateLineColor}

\newcommand{\StateFillCol}{\StateFillColor}
\newcommand{\StateFillSta}{\StateFillStatus} 
\newcommand{\StateLabelSca}{1}
\newcommand{\StateLabelCol}{\StateLabelColor}
\newcommand{\StateDimen}{outer}
\newcommand{\StateDblDimen}{middle}
\newcommand{\VCIFflag}{2}
\newcommand{\PlainState}%
  {\renewcommand{\VCIFflag}{0}}
\newcommand{\FullState}%
  {\renewcommand{\VCIFflag}{2}}

\newif\ifVCShowState

\newcommand{\ShowState}{\VCShowStatetrue}
\ShowState 
\newpsstyle{VaucStateStyle}{framesep=0pt,%
         linewidth=\StateLineWid,linecolor=\StateLineCol,%
         linestyle=\StateLineSty,doubleline=false,%
         fillcolor=\StateFillCol,fillstyle=\StateFillSta,%
         border=0pt,dimen=\StateDimen,%
         cornersize=relative,framearc=1,framesep=0pt}
\newpsstyle{VaucStateDblStyle}{framesep=0pt,%
         linewidth=\StateLineDblCoef\StateLineWid,linecolor=\StateLineCol,%
         linestyle=\StateLineSty,doubleline=true,doublesep=\StateLineDblSep\StateLineWid,%
         fillcolor=\StateFillCol,fillstyle=\StateFillSta,%
         border=0pt,dimen=\StateDblDimen,%
         cornersize=relative,framearc=1,framesep=0pt}
\newpsstyle{VaucHiddenStateStyle}{framesep=0pt,%
         linewidth=\StateLineWid,linecolor=\StateLineCol,%
         linestyle=none,%
         fillcolor=\StateFillCol,fillstyle=none,%
         border=0pt,dimen=outer,%
         cornersize=relative,framearc=1,framesep=0pt}
\newcommand{\StateStyle}{%
   \ifVCShowState%
         \ifStateLineDbl\psset{style=VaucStateDblStyle}\else\psset{style=VaucStateStyle}\fi%
   \else\psset{style=VaucHiddenStateStyle}\fi}
\newcommand{\VaucStateRBLabel}[1]{%
    \textcolor{\StateLabelCol}{\scalebox{\StateLabelSca}{\scalebox{\StateLabelScale}{\rput[B]{0}(0,\VertShift){$ #1 $}}}}}%
\newcommand{\VaucStateLabel}[1]%
    {\ifVCShowState%
        \ifVCRigidLabel%
           \ifVCStateLabelBaseLine%
                 \textcolor{\StateLabelCol}{\scalebox{\StateLabelSca}{\scalebox{\StateLabelScale}{\rput[B]{*0}(0,\VertShift){$ #1 $}}}}%
           \else
                 \textcolor{\StateLabelCol}{\scalebox{\StateLabelSca}{\scalebox{\StateLabelScale}{\rput{*0}(0,0){$ #1 $}}}}%
           \fi
        \else
                 \textcolor{\StateLabelCol}{\scalebox{\StateLabelSca}{\scalebox{\StateLabelScale}{$ #1 $}}}%
        \fi
     \else%
                 \textcolor{white}{\scalebox{\StateLabelSca}{\scalebox{\StateLabelScale}{$ #1 $}}}%
     \fi}
\newcommand{\VCPutStateLabel}[2]%
    {\rput#1{\scalebox{\StateLabelSca}{\scalebox{\StateLabelScale}{$ #2 $}}}}%
\newcommand{\ChgStateLineStyle}[1]{\renewcommand{\StateLineSty}{#1}}
\newcommand{\RstStateLineStyle}{\ChgStateLineStyle{\StateLineStyle}}
\newcommand{\SetStateLineStyle}[1]%
   {\renewcommand{\StateLineStyle}{#1}\RstStateLineStyle}%

\newcommand{\ChgStateLineWidth}[1]{\setlength{\StateLineWid}{#1\StateLineWidth}}%
\newcommand{\RstStateLineWidth}{\ChgStateLineWidth{1}}%
\newcommand{\SetStateLineWidth}[1]
   {\setlength{\StateLineWidth}{#1}\RstStateLineWidth}
\newcommand{\ChgStateLineColor}[1]{\renewcommand{\StateLineCol}{#1}}
\newcommand{\RstStateLineColor}{\ChgStateLineColor{\StateLineColor}}
\newcommand{\SetStateLineColor}[1]%
   {\renewcommand{\StateLineColor}{#1}\RstStateLineColor}
\newcommand{\ChgStateFillStatus}[1]{\renewcommand{\StateFillSta}{#1}}
\newcommand{\RstStateFillStatus}{\ChgStateFillStatus{\StateFillStatus}}
\newcommand{\SetStateFillStatus}[1]%
    {\renewcommand{\StateFillStatus}{#1}\RstStateFillStatus}
\newcommand{\ChgStateFillColor}[1]{\renewcommand{\StateFillCol}{#1}}
\newcommand{\RstStateFillColor}{\ChgStateFillColor{\StateFillColor}}
\newcommand{\SetStateFillColor}[1]%
    {\renewcommand{\StateFillColor}{#1}\RstStateFillColor}%
\newcommand{\ChgStateLabelColor}[1]{\renewcommand{\StateLabelCol}{#1}}
\newcommand{\RstStateLabelColor}{\ChgStateLabelColor{\StateLabelColor}}
\newcommand{\SetStateLabelColor}[1]%
    {\renewcommand{\StateLabelColor}{#1}\RstStateLabelColor}
\newcommand{\ChgStateLabelScale}[1]{\renewcommand{\StateLabelSca}{#1}}
\newcommand{\RstStateLabelScale}{\ChgStateLabelScale{1}}
\newcommand{\SetStateLabelScale}[1]%
   {\renewcommand{\StateLabelScale}{#1}\RstStateLabelScale}

\newcommand{\RstState}{%
   \RstStateLineStyle\RstStateLineWidth%
   \RstStateLineColor%
   \RstStateFillStatus\RstStateFillColor%
   \RstStateLabelColor\RstStateLabelScale}%
\newcommand{\DimState}{%
    \ChgStateLineStyle{\DimStateLineStyle}%
    \ChgStateLineWidth{\DimStateLineCoef}%
    \ChgStateLineColor{\DimStateLineColor}%
        \ChgStateFillColor{\DimStateFillColor}%
    \ChgStateLabelColor{\DimStateLabelColor}}%
\newlength{\StateDiam}
\newlength{\VaucAOS}\newlength{\VaucAOSdiag}
\newcommand{\StateSizeFlag}{}
\newcommand{\SetAOS}{%
   \setlength{\VaucAOS}{\ArrowOnStateCoef\StateDiam}%
   \setlength{\VaucAOSdiag}{\SQRTwo\VaucAOS}}
\newlength{\VariableStateIntDiam}
\newlength{\VariableStateWidth}
\newlength{\VariableStateITPos}
\newcommand{\SetStateIntDiam}{%
   \setlength{\VariableStateIntDiam}{\StateDiam}%
   \addtolength{\VariableStateIntDiam}{-2\StateLineWid}%
}%
\newcommand{\LoopSize}{}\newcommand{\LoopSi}{}
\newcommand{\LoopVarSize}{}\newcommand{\LoopVarSi}{}
\newcommand{\CLoopSize}{}\newcommand{\CLoopSi}{}
\newcommand{\ChgLoopSize}[1]{\renewcommand{\LoopSi}{#1}}
\newcommand{\RstLoopSize}{\ChgLoopSize{\LoopSize}}
\newcommand{\SetLoopSize}[1]%
   {\renewcommand{\LoopSize}{#1}\RstLoopSize}
\newcommand{\ChgCLoopSize}[1]{\renewcommand{\CLoopSi}{#1}}
\newcommand{\RstCLoopSize}{\ChgCLoopSize{\CLoopSize}}
\newcommand{\SetCLoopSize}[1]%
   {\renewcommand{\CLoopSize}{#1}\RstCLoopSize}
\newcommand{\ChgLoopVarSize}[1]{\renewcommand{\LoopVarSi}{#1}}
\newcommand{\RstLoopVarSize}{\ChgLoopVarSize{\LoopVarSize}}
\newcommand{\SetLoopVarSize}[1]%
   {\renewcommand{\LoopVarSize}{#1}\RstLoopVarSize}
\newcommand{\SetStateDiam}[4]{%
   \setlength{\StateDiam}{#1}%
   \renewcommand{\ArrowOnStateCoef}{#2}%
   \SetLoopSize{#3}%
   \SetLoopVarSize{#3}%
   \SetCLoopSize{#4}%
   \SetAOS\SetStateIntDiam}
\newcommand{\FixStateDiameter}[1]
   {\setlength{\StateDiam}{#1}\SetStateIntDiam \SetAOS}
\newcommand{\MediumState}%
   {\SetStateDiam{\MediumStateDiameter}{\ArrowOnMediumState}%
         {\LoopOnMediumState}{\CLoopOnMediumState}%
                  \renewcommand{\StateSizeFlag}{0}}
\newcommand{\SmallState}%
   {\SetStateDiam{\SmallStateDiameter}{\ArrowOnSmallState}%
         {\LoopOnSmallState}{\CLoopOnSmallState}%
                  \renewcommand{\StateSizeFlag}{1}}
\newcommand{\LargeState}%
   {\SetStateDiam{\LargeStateDiameter}{\ArrowOnLargeState}%
         {\LoopOnLargeState}{\CLoopOnLargeState}%
                  \renewcommand{\StateSizeFlag}{2}}
\newcommand{\RstStateSize}%
  {\ifthenelse{\equal{\StateSizeFlag}{0}}%
              {\MediumState}%
              {\ifthenelse{\equal{\StateSizeFlag}{1}}%
                              {\SmallState}{\LargeState}}}%
\MediumState
\newcommand{\VaucState}[3][{}]%
   {\rput#2{%
      \Cnode[radius=.5\StateDiam](0,0){#3}%
          \ifVCShowState%
      \nput[labelsep=-.5\StateDiam]{0}{#3}%
        {\makebox[0pt]{\VaucStateLabel{#1}}}%
      \fi%
      \ifthenelse{\equal{\VCIFflag}{0}}{}{%
        \pnode(-\VaucAOS,0){#3w}\pnode(\VaucAOS,0){#3e}%
        \pnode(0,\VaucAOS){#3n}\pnode(0,-\VaucAOS){#3s}%
           \ifthenelse{\equal{\VCIFflag}{1}}{}{%
          \pnode(-\VaucAOSdiag,\VaucAOSdiag){#3nw}%
           \pnode(\VaucAOSdiag,\VaucAOSdiag){#3ne}%
           \pnode(-\VaucAOSdiag,-\VaucAOSdiag){#3sw}%
           \pnode(\VaucAOSdiag,-\VaucAOSdiag){#3se}%
                   }%
            }%
     }%
}
\newcommand{\State}[3][{}]{\StateStyle\VaucState[#1]{#2}{#3}}
\newcommand{\FinalState}[3][{}]%
   {\psset{style=VaucStateDblStyle}\VaucState[#1]{#2}{#3}}
\newcommand{\VSState}[2]%
    {\renewcommand{\ArrowOnStateCoef}{\ArrowOnVerySmallState}%
         \FixStateDiameter{\VerySmallStateDiameter}%
     \ChgStateLineWidth{\VSStateLineCoef}%
         \State{#1}{#2}%
         \RstStateLineWidth\RstStateSize}
\newlength{\ExtraSpace}
\newcommand{\StateVar}[3][]%
 {\StateStyle %
  \settowidth{\VariableStateWidth}{\scalebox{\StateLabelSca}{\scalebox{\StateLabelScale}{$#1$}}}%
  \addtolength{\VariableStateWidth}{\ExtraSpace}
  \ifthenelse{\lengthtest{\VariableStateWidth < \VariableStateIntDiam}}%
        {\setlength{\VariableStateWidth}{\VariableStateIntDiam}}{}%
  \setlength{\VariableStateITPos}{\ArrowOnStateCoef\StateDiam}%
  \addtolength{\VariableStateITPos}{0.5\VariableStateWidth}%
  \addtolength{\VariableStateITPos}{-0.5\StateDiam}%
  \rput#2{\pnode(\VariableStateITPos,0){#3e}%
          \pnode(-\VariableStateITPos,0){#3w}%
          \pnode(0,\ArrowOnStateCoef\StateDiam){#3n}%
          \pnode(0,-\ArrowOnStateCoef\StateDiam){#3s}}%
  \rput#2{\rnode{#3}{\psframebox{\protect\rule[-.5\VariableStateIntDiam]{0pt}{\VariableStateIntDiam}\protect\rule{\VariableStateWidth}{0pt}}}}
  \rput#2{\VaucStateRBLabel{#1}}%
}%

\newlength{\EdgeLineWid}
\newcommand{\EdgeLineSty}{\EdgeLineStyle}
\newcommand{\EdgeLineCol}{\EdgeLineColor}
\newcommand{\EdgeLabelSca}{1}
\newcommand{\EdgeLabelCol}{\EdgeLabelColor}
\newlength{\EdgeArrowSZDim}
\newcommand{\EdgeArrowSZNum}{\EdgeArrowLengthCoef}
\newcommand{\EdgeArrowSty}{\EdgeArrowStyle}
\newcommand{\EdgeArrowIns}{\EdgeArrowInset}
\newlength{\EdgeLineBord}
\newlength{\ZZSiZ}
\newcommand{\ZZLineWid}{\ZZLineWidth}
\newlength{\EdgeOff}
\newcommand{\VaucArcAng}{\VaucArcAngle}
\newcommand{\VaucLArcAng}{\VaucLArcAngle}
\newlength{\VaucArcOff}
\newcommand{\VaucArcCurv}{\VaucArcCurvature}
\newcommand{\VaucLArcCurv}{\VaucLArcCurvature}
\newcommand{\LoopAng}{\LoopAngle}
\newcommand{\CLoopAng}{\CLoopAngle}
\newcommand{\LoopVarAng}{\LoopVarAngle}
\newlength{\LoopOff}
\newlength{\LoopVarOff}
\newlength{\TransLabelSP}
\newcommand{\EdgeLabelPos}{\EdgeLabelPosit}
\newcommand{\ArcLabelPos}{\ArcLabelPosit}
\newcommand{\LArcLabelPos}{\LArcLabelPosit}
\newcommand{\LoopLabelPos}{\LoopLabelPosit}
\newcommand{\CLoopLabelPos}{\CLoopLabelPosit}
\newcommand{\InitStateLabelPos}{\InitStateLabelPosit}
\newcommand{\FinalStateLabelPos}{\FinalStateLabelPosit}
\newpsstyle{VaucEdgeStyle}%
    {arrows=\EdgeArrowSty,arrowsize=\EdgeArrowSZDim,arrowlength=\EdgeArrowSZNum,%
         arrowinset=\EdgeArrowIns,%
     linewidth=\EdgeLineWid,linecolor=\EdgeLineCol,linestyle=\EdgeLineSty,%
     doubleline=false,%
         bordercolor=\EdgeLineBorderColor,border=\EdgeLineBord,%
     fillstyle=none,offset=\EdgeOff,%
     labelsep=\TransLabelSP,nodesep=0pt}
\newpsstyle{VaucEdgeDblStyle}%
    {arrows=\EdgeArrowSty,arrowsize=\EdgeArrowSZDim,arrowlength=\EdgeArrowSZNum,%
         arrowinset=\EdgeArrowIns,%
     linewidth=\EdgeLineDblCoef\EdgeLineWid,linecolor=\EdgeLineCol,linestyle=\EdgeLineSty,%
     doubleline=true,doublesep=\EdgeLineDblSep\EdgeLineWid,%
         bordercolor=\EdgeLineBorderColor,border=\EdgeLineBord,%
     fillstyle=none,offset=\EdgeOff,%
     labelsep=\TransLabelSP,nodesep=0pt}
\newpsstyle{VaucArcR}{ncurv=\VaucArcCurv,arcangle=-\VaucArcAng,%
     labelsep=\TransLabelSP,offset=-\VaucArcOff}
\newpsstyle{VaucArcL}{ncurv=\VaucArcCurv,arcangle=\VaucArcAng,%
     labelsep=\TransLabelSP,offset=\VaucArcOff}
\newpsstyle{VaucLArcR}{ncurv=\VaucLArcCurv,arcangle=-\VaucLArcAng,%
     labelsep=\TransLabelSP,offset=-\VaucArcOff}
\newpsstyle{VaucLArcL}{ncurv=\VaucLArcCurv,arcangle=\VaucLArcAng,%
     labelsep=\TransLabelSP,offset=\VaucArcOff}
\newpsstyle{VaucZigzagStyle}%
   {linewidth=\ZZLineWid\EdgeLineWid,%
    labelsep=\TransLabelSP,nodesep=0pt,%
    coilwidth=1.2\ZZSiZ,coilarmA=0.1\ZZSiZ,%
    coilarmB=0.3\ZZSiZ,coilheight=\ZZShape,linearc=1.6pt}
\newcommand{\EdgeStyle}{\ifEdgeLineDbl\psset{style=VaucEdgeDblStyle}%
        \else\psset{style=VaucEdgeStyle}\fi}
\newcommand{\ZigzagStyle}%
   {\addtolength{\TransLabelSP}{\TransLabelZZCoef\ZZSiZ}%
    \psset{style=VaucZigzagStyle}%
        \addtolength{\TransLabelSP}{-\TransLabelZZCoef\ZZSiZ}%
        }
\newcommand{\ChgEdgeOffset}[1]{\setlength{\EdgeOff}{#1}}
\newcommand{\RstEdgeOffset}{\ChgEdgeOffset{\EdgeOffset}}
\newcommand{\SetEdgeOffset}[1]%
   {\setlength{\EdgeOffset}{#1}\RstEdgeOffset}

\newcommand{\ChgArcAngle}[1]{\renewcommand{\VaucArcAng}{#1}}
\newcommand{\RstArcAngle}{\ChgArcAngle{\VaucArcAngle}}
\newcommand{\SetArcAngle}[1]%
   {\renewcommand{\VaucArcAngle}{#1}\RstArcAngle}
\newcommand{\ChgLArcAngle}[1]{\renewcommand{\VaucLArcAng}{#1}}
\newcommand{\RstLArcAngle}{\ChgLArcAngle{\VaucLArcAngle}}
\newcommand{\SetLArcAngle}[1]%
   {\renewcommand{\VaucLArcAngle}{#1}\RstLArcAngle}
\newcommand{\ChgArcCurvature}[1]{\renewcommand{\VaucArcCurv}{#1}}
\newcommand{\RstArcCurvature}{\ChgArcCurvature{\VaucArcCurvature}}
\newcommand{\SetArcCurvature}[1]%
   {\renewcommand{\VaucArcCurvature}{#1}\RstArcCurvature}
\newcommand{\ChgLArcCurvature}[1]{\renewcommand{\VaucLArcCurv}{#1}}
\newcommand{\RstLArcCurvature}{\ChgLArcCurvature{\VaucLArcCurvature}}
\newcommand{\SetLArcCurvature}[1]%
   {\renewcommand{\VaucLArcCurvature}{#1}\RstLArcCurvature}

\newcommand{\RstArcOffset}{\setlength{\VaucArcOff}{\VaucArcOffset}}
\newcommand{\SetArcOffset}[1]%
   {\renewcommand{\VaucArcOffset}{#1}\RstArcOffset}

\newcommand{\RstLoopOffset}{\setlength{\LoopOff}{\LoopOffset}}
\newcommand{\SetLoopOffset}[1]%
   {\renewcommand{\LoopOffset}{#1}\RstLoopOffset}
\newcommand{\ChgLoopAngle}[1]{\renewcommand{\LoopAng}{#1}}
\newcommand{\RstLoopAngle}{\ChgLoopAngle{\LoopAngle}}
\newcommand{\SetLoopAngle}[1]%
   {\renewcommand{\LoopAngle}{#1}\RstLoopAngle}
\newcommand{\ChgCLoopAngle}[1]{\renewcommand{\CLoopAng}{#1}}
\newcommand{\RstCLoopAngle}{\ChgCLoopAngle{\CLoopAngle}}
\newcommand{\SetCLoopAngle}[1]%
   {\renewcommand{\CLoopAngle}{#1}\RstCLoopAngle}
\newcommand{\ChgEdgeLineColor}[1]{\renewcommand{\EdgeLineCol}{#1}}
\newcommand{\RstEdgeLineColor}{\ChgEdgeLineColor{\EdgeLineColor}}
\newcommand{\SetEdgeLineColor}[1]%
   {\renewcommand{\EdgeLineColor}{#1}\RstEdgeLineColor}
\newcommand{\ChgEdgeLineStyle}[1]{\renewcommand{\EdgeLineSty}{#1}}  
\newcommand{\RstEdgeLineStyle}{\ChgEdgeLineStyle{\EdgeLineStyle}}
\newcommand{\SetEdgeLineStyle}[1]%
   {\renewcommand{\EdgeLineStyle}{#1}\RstEdgeLineStyle}
\newcommand{\ChgEdgeLineWidth}[1]
   {\setlength{\EdgeLineWid}{#1\EdgeLineWidth}}
\newcommand{\RstEdgeLineWidth}{\ChgEdgeLineWidth{1}}
\newcommand{\SetEdgeLineWidth}[1]
   {\setlength{\EdgeLineWidth}{#1}\RstEdgeLineWidth}
\newcommand{\EdgeLineDouble}%
        {\EdgeLineDbltrue%
    \ChgEdgeArrowWidth{\EdgeDblArrowWidth}
    \ChgEdgeArrowLengthCoef{\EdgeDblArrowLengthCoef}}
\newcommand{\EdgeLineSimple}%
   {\EdgeLineDblfalse \RstEdgeArrowWidth \RstEdgeArrowLengthCoef}
\newcommand{\ChgEdgeLabelColor}[1]{\renewcommand{\EdgeLabelCol}{#1}}
\newcommand{\RstEdgeLabelColor}{\ChgEdgeLabelColor{\EdgeLabelColor}}
\newcommand{\SetEdgeLabelColor}[1]%
   {\renewcommand{\EdgeLabelColor}{#1}\RstEdgeLabelColor}
\newcommand{\ChgEdgeLabelScale}[1]{\renewcommand{\EdgeLabelSca}{#1}}
\newcommand{\RstEdgeLabelScale}{\ChgEdgeLabelScale{1}}
\newcommand{\SetEdgeLabelScale}[1]%
   {\renewcommand{\EdgeLabelScale}{#1}\RstEdgeLabelScale}
\newcommand{\ChgEdgeArrowStyle}[1]{\renewcommand{\EdgeArrowSty}{#1}}
\newcommand{\RstEdgeArrowStyle}{\ChgEdgeArrowStyle{\EdgeArrowStyle}}
\newcommand{\SetEdgeArrowStyle}[1]%
   {\renewcommand{\EdgeArrowStyle}{#1}\RstEdgeArrowStyle}
\newcommand{\ChgEdgeArrowWidth}[1]%
   {\setlength{\EdgeArrowSZDim}{#1}} 
\newcommand{\RstEdgeArrowWidth}{\ChgEdgeArrowWidth{\EdgeArrowWidth}}
\newcommand{\SetEdgeArrowWidth}[1]%
   {\setlength{\EdgeArrowWidth}{#1} \RstEdgeArrowWidth}
\newcommand{\ChgEdgeArrowLengthCoef}[1]{\renewcommand{\EdgeArrowSZNum}{#1}}
\newcommand{\RstEdgeArrowLengthCoef}{\ChgEdgeArrowLengthCoef{\EdgeArrowLengthCoef}}
\newcommand{\SetEdgeArrowLengthCoef}[1]%
   {\renewcommand{\EdgeArrowLengthCoef}{#1}\RstEdgeArrowLengthCoef}
\newcommand{\ChgEdgeArrowInsetCoef}[1]{\renewcommand{\EdgeArrowIns}{#1}}
\newcommand{\RstEdgeArrowInsetCoef}{\ChgEdgeArrowInsetCoef{\EdgeArrowInset}}
\newcommand{\SetEdgeArrowInsetCoef}[1]%
   {\renewcommand{\EdgeArrowInset}{#1}\RstEdgeArrowInsetCoef}
\newcommand{\ReverseArrow}%
   {\ChgEdgeArrowStyle{\EdgeRevArrowStyle}%
    \renewcommand{\EdgeLabelPos}{\EdgeLabelRevPosit}%
    \renewcommand{\ArcLabelPos}{\ArcLabelRevPosit}%
    \renewcommand{\LArcLabelPos}{\LArcLabelRevPosit}%
    \renewcommand{\LoopLabelPos}{\LoopLabelRevPosit}%
    \renewcommand{\CLoopLabelPos}{\CLoopLabelRevPosit}%
    \renewcommand{\InitStateLabelPos}{\InitStateLabelRevPosit}%
    \renewcommand{\FinalStateLabelPos}{\FinalStateLabelRevPosit}}
\newcommand{\StraightArrow}%
   {\ChgEdgeArrowStyle{\EdgeArrowStyle}%
    \renewcommand{\EdgeLabelPos}{\EdgeLabelPosit}%
    \renewcommand{\ArcLabelPos}{\ArcLabelPosit}%
    \renewcommand{\LArcLabelPos}{\LArcLabelPosit}%
    \renewcommand{\LoopLabelPos}{\LoopLabelPosit}%
    \renewcommand{\CLoopLabelPos}{\CLoopLabelPosit}%
    \renewcommand{\InitStateLabelPos}{\InitStateLabelPosit}%
    \renewcommand{\FinalStateLabelPos}{\FinalStateLabelPosit}}
\newcommand{\FixEdgeLineDouble}[2]{%
    \renewcommand{\EdgeLineDblCoef}{#1}%
    \renewcommand{\EdgeLineDblSep}{#2}}

\newcommand{\EdgeBorder}%
  {\setlength{\EdgeLineBord}{\EdgeLineBorderCoef\EdgeLineWid}}

\newcommand{\DimEdge}{%
    \ChgEdgeLineStyle{\DimEdgeLineStyle}%
    \ChgEdgeLineWidth{\DimEdgeLineCoef}%
    \ChgEdgeLineColor{\DimEdgeLineColor}%
        \ChgEdgeLabelColor{\DimEdgeLabelColor}}

\newcommand{\ChgZZLineWidth}[1]{\renewcommand{\ZZLineWid}{#1}}
\newcommand{\RstZZLineWidth}{\ChgZZLineWidth{\ZZLineWidth}}
\newcommand{\SetZZLineWidth}[1]%
   {\renewcommand{\ZZLineWidth}{#1}\RstZZLineWidth}
\newcommand{\VaucEdgeLabel}[1]{%
        \textcolor{\EdgeLabelCol}{\scalebox{\EdgeLabelSca}{\scalebox{\EdgeLabelScale}{$ #1 $}}}}
\newcommand{\RstEdge}{%
   \RstEdgeOffset\RstArcAngle\RstLArcAngle%
   \RstArcCurvature\RstLArcCurvature%
   \RstArcOffset\RstLoopOffset\RstLoopSize%
   \RstEdgeLineColor\RstEdgeLineStyle\RstEdgeLineWidth\EdgeLineSimple%
   \StraightArrow
   \RstEdgeLabelScale\RstEdgeLabelColor}

\newcommand{\Initial}[2][\InitialDir]{\EdgeStyle\ncline{#2#1}{#2}}
\newcommand{\Final}[2][\FinalDir]{\EdgeStyle\ncline{#2}{#2#1}}
\newcommand{\InitialL}[4][{\InitStateLabelPos}]%
   {\EdgeStyle\ncline{#3#2}{#3}\naput[npos=#1]{\VaucEdgeLabel{#4}}}
\newcommand{\InitialR}[4][{\InitStateLabelPos}]%
   {\EdgeStyle\ncline{#3#2}{#3}\nbput[npos=#1]{\VaucEdgeLabel{#4}}}
\newcommand{\FinalL}[4][{\FinalStateLabelPos}]%
   {\EdgeStyle\ncline{#3}{#3#2}\naput[npos=#1]{\VaucEdgeLabel{#4}}}
\newcommand{\FinalR}[4][{\FinalStateLabelPos}]%
   {\EdgeStyle\ncline{#3}{#3#2}\nbput[npos=#1]{\VaucEdgeLabel{#4}}}
\newcommand{\EdgeL}[4][{\EdgeLabelPos}]%
   {\EdgeStyle \ncline{#2}{#3} \naput[npos=#1]{\VaucEdgeLabel{#4}}}
\newcommand{\EdgeR}[4][{\EdgeLabelPos}]%
   {\EdgeStyle \ncline{#2}{#3} \nbput[npos=#1]{\VaucEdgeLabel{#4}}}
\newcommand{\ArcL}[4][{\ArcLabelPos}]%
   {\EdgeStyle \psset{style=VaucArcL}%
    \ncarc{#2}{#3} \naput[npos=#1]{\VaucEdgeLabel{#4}}}
\newcommand{\ArcR}[4][{\ArcLabelPos}]%
   {\EdgeStyle \psset{style=VaucArcR}%
    \ncarc{#2}{#3} \nbput[npos=#1]{\VaucEdgeLabel{#4}}}
\newcommand{\LArcL}[4][{\LArcLabelPos}]%
   {\EdgeStyle \psset{style=VaucLArcL}%
    \ncarc{#2}{#3} \naput[npos=#1]{\VaucEdgeLabel{#4}}}
\newcommand{\LArcR}[4][{\LArcLabelPos}]%
   {\EdgeStyle \psset{style=VaucLArcR}%
    \ncarc{#2}{#3} \nbput[npos=#1]{\VaucEdgeLabel{#4}}}
\newcounter{anglea}\newcounter{angleb}
\newcommand{\LoopXR}[7]%
   {{\setcounter{anglea}{#2-#4}}%
    {\setcounter{angleb}{#2+#4}}%
    {\EdgeStyle \psset{angleA=\theanglea,angleB=\theangleb,offset=#5,ncurv=#6}%
    \nccurve{#3}{#3} \nbput[npos=#1]{\VaucEdgeLabel{#7}}}}
\newcommand{\LoopXL}[7]%
   {{\setcounter{anglea}{#2+#4}}%
    {\setcounter{angleb}{#2-#4}}%
    {\EdgeStyle \psset{angleA=\theanglea,angleB=\theangleb,offset=-#5,ncurv=#6}%
    \nccurve{#3}{#3} \naput[npos=#1]{\VaucEdgeLabel{#7}}}}
\newcommand{\LoopR}[4][{\LoopLabelPos}]%
   {\LoopXR{#1}{#2}{#3}{\LoopAng}{\LoopOff}{\LoopSi}{#4}}
\newcommand{\LoopL}[4][{\LoopLabelPos}]%
   {\LoopXL{#1}{#2}{#3}{\LoopAng}{\LoopOff}{\LoopSi}{#4}}
\newcommand{\CLoopR}[4][{\CLoopLabelPos}]%
   {\LoopXR{#1}{#2}{#3}{\CLoopAng}{\LoopOff}{\LoopSi}{#4}}
\newcommand{\CLoopL}[4][{\CLoopLabelPos}]%
   {\LoopXL{#1}{#2}{#3}{\CLoopAng}{\LoopOff}{\LoopSi}{#4}}
\newcommand{\LoopVarR}[4][{\LoopLabelPos}]%
   {\LoopXR{#1}{#2}{#3}{\LoopVarAng}{\LoopVarOff}{\LoopVarSi}{#4}}
\newcommand{\LoopVarL}[4][{\LoopLabelPos}]%
   {\LoopXL{#1}{#2}{#3}{\LoopVarAng}{\LoopVarOff}{\LoopVarSi}{#4}}
\newcommand{\LoopW}[3][{\LoopLabelPos}]{\LoopR[#1]{180}{#2}{#3}}
\newcommand{\LoopE}[3][{\LoopLabelPos}]{\LoopL[#1]{0}{#2}{#3}}
\newcommand{\LoopN}[3][{\LoopLabelPos}]{\LoopL[#1]{90}{#2}{#3}}
\newcommand{\LoopS}[3][{\LoopLabelPos}]{\LoopR[#1]{-90}{#2}{#3}}
\newcommand{\LoopNW}[3][{\LoopLabelPos}]{\LoopR[#1]{135}{#2}{#3}}
\newcommand{\LoopNE}[3][{\LoopLabelPos}]{\LoopL[#1]{45}{#2}{#3}}
\newcommand{\LoopSW}[3][{\LoopLabelPos}]{\LoopL[#1]{-135}{#2}{#3}}
\newcommand{\LoopSE}[3][{\LoopLabelPos}]{\LoopR[#1]{-45}{#2}{#3}}

\newcommand{\ZZEdge}[2]%
   {\EdgeStyle\ZigzagStyle\nczigzag{#1}{#2}}%
\newcommand{\ZZEdgeL}[4][{\EdgeLabelRevPosit}]%
   {\EdgeStyle\ZigzagStyle\nczigzag{#2}{#3}%
    \naput[npos=#1]{\VaucEdgeLabel{#4}}}
\newcommand{\ZZEdgeR}[4][{\EdgeLabelRevPosit}]%
   {\EdgeStyle\ZigzagStyle\nczigzag{#2}{#3}%
    \nbput[npos=#1]{\VaucEdgeLabel{#4}}}

\newcommand{\VArcL}[5][{\ArcLabelPos}]%
   {\EdgeStyle \psset{style=VaucLArcL}%
    \ncarc[#2]{#3}{#4} \naput[npos=#1]{\VaucEdgeLabel{#5}}}
\newcommand{\VArcR}[5][{\ArcLabelPos}]%
   {\EdgeStyle \psset{style=VaucLArcR}%
    \ncarc[#2]{#3}{#4} \nbput[npos=#1]{\VaucEdgeLabel{#5}}}
\newcommand{\VCurveL}[5][{\ArcLabelPos}]%
   {\EdgeStyle \psset{angleA=0,angleB=180,ncurv=1}%
    \nccurve[#2]{#3}{#4} \naput[npos=#1]{\VaucEdgeLabel{#5}}}
\newcommand{\VCurveR}[5][{\ArcLabelPos}]%
   {\EdgeStyle \psset{angleA=0,angleB=0,ncurv=1}%
    \nccurve[#2]{#3}{#4} \nbput[npos=#1]{\VaucEdgeLabel{#5}}}
\newcommand{\LabelL}[2][{\EdgeLabelPos}]%
   {\naput[npos=#1]{\VaucEdgeLabel{#2}}}
\newcommand{\LabelR}[2][{\EdgeLabelPos}]%
   {\nbput[npos=#1]{\VaucEdgeLabel{#2}}}
\def\math#1{\ifmmode #1\else \mbox{$#1$}\xspace \fi}
\def\mathnosp#1{\ifmmode #1\else \mbox{$#1$}\fi}
\newcommand{\Pfrak}{\mathfrak{P}}
\newcommand{\jsPart}[1]{\math{\Pfrak(#1)}}
\newcommand{\jspart}[1]{\jsPart{#1}} 

\newcommand{\jsPartK}[2]{\math{\Pfrak_{(#2)}(#1)}}

\renewcommand{\max}{\math{\operatornamewithlimits{\mathsf{max}}}}

\newcommand{\lmn}{\math{(\lambda , \mu , \nu )}}
\newcommand{\x}{\! \times \!}

\newcommand{\matmul}{\mathbin{\cdot}}

\newcommand{\jsStar}[1]{\math{{#1}^{*}}}
\newcommand{\Tran}[1]{\bigl ( #1\bigr )}
\newcommand{\Be}{\jsStar{B}}
\newcommand{\Rat}{\mathrm{Rat}\,}
\newcommand{\BeBe}{\math{\Be \x \Be }}
\newcommand{\CompAuto}[1]{\math{\pmb{|}{#1}\pmb{|}}}
\newcommand{\compA}{\CompAuto{\Ac}}

\newcommand{\comput}[1]{\xrightarrow{ #1 }}
\newcommand{\computaut}[2]{\underset{#2}{\comput{#1}}}
\newcommand{\varqed}{\hfill \ensuremath{\square}}
\theoremstyle{definition}
\newtheorem{property}[thm]{Property}

\newsavebox{\InterSymbolSpace}
\savebox{\InterSymbolSpace}{\hspace{0.125em}}
\newsavebox{\SideFormulaSpace}
\savebox{\SideFormulaSpace}{\hspace{0.2em}}
\newcommand{\msp}{\usebox{\SideFormulaSpace}} 
\newcommand{\xmd}{\usebox{\InterSymbolSpace}} 
\newcommand{\ETAbar}[1]{\overline{\rule{0pt}{1.5ex}\smash{#1}}}
\newcommand{\Ac}{\mathcal{A}}
\newcommand{\Bc}{\mathcal{B}}
\newcommand{\Cc}{\mathcal{C}}
\newcommand{\Dc}{\mathcal{D}}
\newcommand{\Oc}{\mathcal{O}}
\newcommand{\Sc}{\mathcal{S}}
\newcommand{\Tc}{\mathcal{T}}
\newcommand{\Uc}{\mathcal{U}}
\newcommand{\Vc}{\mathcal{V}}
\newcommand{\Wc}{\mathcal{W}}
\newcommand{\Zc}{\mathcal{Z}}
\newcommand{\Nmbb}{\mathbb{N}}
\newcommand{\Bbar}{\ETAbar{B}}
\newcommand{\ubar}{\ETAbar{u}}

\newcommand{\xbar}{\ETAbar{x}}

\newcommand{\lda}{\cdot}
\newcommand{\smtrans}{\xi} 

\newcommand{\ldzero}{\mathbf{0}}

\newcommand{\Nvec}[1]{\ensuremath{\mathsf{#1}}}
\newcommand{\WSvec}[1]{\ensuremath{\mathsf{#1}}}

\newcommand{\computvec}[1]{\ensuremath{\mathbf{#1}}}

\newcommand{\card}[1]{\ensuremath{\mathrm{card} \left ( #1 \right) }}

\newcommand{\monus}{\ensuremath{-}}

\newcommand{\ew}{\ensuremath{1}} 
\newcommand{\ewfg}{\ensuremath{\mathbf{1}}} 

\newcommand{\inD}{\rho}

\newcommand{\zv}{\ensuremath{\vec{0}}}

\newcommand{\lag}[2]{\ensuremath{\langle #1, #2 \rangle}}
\newcommand{\lagset}[1]{\ensuremath{\Delta_{#1}}}
\newcommand{\lagN}{\lagset{N}}
\newcommand{\LD}[2]{\ensuremath{\mathrm{LD}(#1, #2)}}
\newcommand{\LDset}[1]{\Delta}

\newcommand{\llex}{\prec}
\newcommand{\FG}[1]{F(#1)} 

\newcommand{\T}{\Tc}     
\newcommand{\Tpsc}{\Uc_N}  
\newcommand{\Tpsi}{\Vc_N}  
\newcommand{\Aka}{\Ac}   
\newcommand{\Asi}{\Bc}   
\newcommand{\Asil}{\Zc}  
\newcommand{\Tck}{\Tc^{k + 1}}     

\newcommand{\dotted}[1] {\psset{dotsep=2.5pt}
    \ChgEdgeLineStyle{dotted}%
    \ChgEdgeLineWidth{2.7}
    \ChgEdgeLabelScale{0.85}
    #1 \RstEdgeLineStyle%
    \RstEdgeLineWidth%
    \RstEdgeLabelScale}

\newcommand{\dashed}[1] {
    \ChgEdgeLineStyle{dashed}%
    \ChgEdgeLineWidth{2.7}
    \ChgEdgeLabelScale{0.85}
    #1 \RstEdgeLineStyle%
    \RstEdgeLineWidth%
    \RstEdgeLabelScale}

\begin{document}

\title{On the decomposition of $k$-valued rational relations}

\author[lab1]{Jacques Sakarovitch}{Jacques Sakarovitch}
\address[lab1]{LTCI, ENST/CNRS, Paris (France)}
\email{sakarovitch@enst.fr}
\author[lab2]{Rodrigo de Souza}{Rodrigo de Souza}
\address[lab2]{ENST, 46, rue Barrault, 75634 Paris Cedex 13 (France)}
\email{rsouza@enst.fr}


\begin{abstract}
  \noindent We give a new, and hopefully more easily understandable,
  structural proof of the decomposition of a $k$-valued transducer
  into $k$ unambiguous functional ones, a result established by
  A. Weber in 1996.
  Our construction is based on a lexicographic ordering of
  computations of automata and on two coverings that can be build by
  means of this ordering.
  The complexity of the construction, measured as the number of states
  of the transducers involved in the decomposition, improves the
  original one by one exponential.
  Moreover, this method allows further generalisation that solves
  the problem of decomposition of rational relations with
  bounded length-degree, which was left open in Weber's paper.
\end{abstract}

\maketitle

\stacsheading{2008}{621-632}{Bordeaux}
\firstpageno{621}

\vspace{-.2cm}

\begin{center}
  \textsc{Extended abstract}
\end{center}

\vspace{-.2cm}

\section{Introduction}
\label{s.intro}

This communication is part of a complete reworking\footnote{A
  financial support of CAPES Foundation (Brazilian government)
  for doctoral studies is gratefully acknowledged by the second
  author (second in the alphabetical order, as in use in the British
  and French encyclopedias --- not in the Lusitanian ones).}
and rewriting of the theory of $k$-valued rational relations and
transducers which puts it in line with the theory of rational
functions ($1$-valued rational relations) and functional transducers
and makes it appear as a natural generalisation of the latter not only
at the level of the results --- as we recall in the next paragraph ---
but also \emph{at the level of proofs}.

It is decidable whether a transducer is
functional (originally due to Schützenberger~\cite{Sch75}); as
a consequence, the equivalence of functional transducers is decidable,
and, above all, every functional transducer is equivalent to an
unambiguous one~\cite{Eil74}.
These results generalise in a remarkable way to bounded
valued rational relations and transducers.
It is decidable whether the image of every word by a given transducer
is bounded (Weber~\cite{Weber1990}), it is decidable whether it is
bounded by a given integer $k$ (Gurari and Ibarra~\cite{GI83}),
every $k$-valued transducer is equivalent to the sum of $k$ functional
(and thus unambiguous) ones (Weber~\cite{Weber96}) and the equivalence
of $k$-valued transducers is decidable (Culik and
Karhum\"aki~\cite{KarhumakiCulik86}).

It is noteworthy that all the results just quoted for functional
transducers are now (if not in the original papers) established
by means of constructions conducted on the transducers themselves
[2,9,11] whereas the corresponding results on $k$-valued transducers
come, in some sense, ``from outside'' and, what is worse, from a
different world for each of them.
Gurari and Ibarra's proof for the
decidability of the $k$-valuedness relies on a reduction to the 
emptiness problem for a class of counter automata, Culik and
Karhum\"aki's one for the decidability of the equivalence appears in the
context of the solution of Ehrenfeucht's conjecture on HDTOL languages, 
and Weber's proof of the decomposition --- which we shall discuss more  
in detail below --- is highly combinatorial and still somewhat detached 
from the transducers.

Our approach for those results are based on constructions which depend
directly on the structure of the automata.
They give back the subject a full coherence and yield systematically
better complexity bounds.
This will be illustrated in this paper with
a new proof of the decomposition
theorem which we restate below as Theorem~\ref{t.weber}.
In~\cite{SakaSouzaLATIN08} we give a new proof for the decidability of
the $k$-valuedness. 

\begin{theorem}[Weber~\cite{Weber96}]
  Every $k$-valued transducer $\Tc$ can be effectively
  decomposed into a sum of $k$ (unambiguous) functional
  transducers.\footnote{By ``decomposed'' we mean that the relation
    realised by $\T$ and the union of the relations realised by the
    $k$ transducers are the same.}
  \label{t.weber}
\end{theorem}

Our proof for Theorem~\ref{t.weber} differs from the original one by
three aspects.
First, Weber's proof is generally considered as very difficulty to
follow, whereas ours is hopefully simpler.
Second, Weber's construction results in $k$ transducers whose number
of states is a double exponential on the number of states of $\T$,
whereas we obtain a decomposition of single exponential size.
Third and finally, our method allows to solve the problem, posed by
Weber, of the decomposition of bounded length-degree rational
relations with a more general statement
(in Weber's question, $\theta$ is the length morphism):

\begin{theorem}
  Let $\tau: A^* \rightarrow B^*$ be a finite image rational relation
  and $\theta: B^* \rightarrow C^*$ a morphism such that the
  composition $\tau \theta$ is $k$-valued.\footnote{We write
    functions and relations using a postfix notation: $x\tau$ is
    the image of $x$ by the relation $\tau$ and thus the composition
    of relations is written by left-to-right concatenation.
    Let us recall that the rational relations
    are closed under composition~\cite{Eil74}.}
  Every transducer $\Sc$ realising $\tau$ can be effectively decomposed
  into $k$ transducers whose compositions with $\theta$ are
  functions.
  \label{t.gen-weber}
\end{theorem}

Our proof makes use twice of the notion of covering of
automata. 
A covering of an automaton\footnote{As we shall define in
Section~\ref{s.prem}, transducers are automata of a certain kind.}
$\Ac$ is an \emph{expansion} of~$\Ac$: a new
automaton $\Bc$ whose states and transitions map to those of~$\Ac$,
preserving adjacency and labels of transitions.
Moreover, the outgoing transitions of every state of $\Bc$
map one-to-one to those of the projection, which implies
a bijection between the successful computations of $\Ac$ and $\Bc$.
Typically, $\Bc$ is larger than $\Ac$, for several states
can have the same image. 
This allows to choose certain subsets of the computations
of $\Ac$ by erasing parts of $\Bc$.

The two coverings we are going to define are based on a lexicographic
ordering on the computations.
This method can be seen as a conceptual generalisation of the one
used by H. Johnson in order to build a lexicographic
selection of deterministic rational 
relations~\cite{Johnson85,Johnson86TCS}.

The first construction, explained
in Section~\ref{s.N-sep}, is what we call the \emph{lag separation
  covering}~$\Tpsc$ of a (real time) transducer $\T$.
It is parameterised by an integer~$N$, and roughly speaking allows to
distinguish between computations with same input and
same output and whose \emph{lag}\footnote{To be defined in the body of
  the paper.} is bounded by~$N$.
If $\T$ is $k$-valued, we show that for a certain~$N$, $\Tpsc$
contains a subtransducer~$\Tpsi$ which is equivalent to~$\T$ and
input-$k$-ambiguous\footnote{%
   When it comes to ambiguity in transducers, we distinguish 
   between \emph{input-ambiguity} (called ambiguity in most 
   of the references) and ambiguity of the transducer (which allows 
   to define ambiguity for relations).}
 (Proposition~\ref{p.N-sep-k-amb}).

The second construction
(Section~\ref{s.lex-cov}) is what we call the {\it multi-skimming
  covering} of an $\Nmbb$-automaton.
It proves the following {\it multi-skimming theorem} for
$\Nmbb$-rational series:
\begin{theorem}
  Let $\Ac$ be a finite $\Nmbb$-automaton with $n$ states
  realising the series $s$.
  There exists an infinite $\Nmbb$-covering $\Bc$ of $\Ac$
  such that for every integer $k > 0$, there exists a finite
  $\mathbb{N}$-quotient $\Bc_k$ of $\Bc$ which satisfies: 
  $\Bc_k$ is an $\Nmbb$-covering of $\Ac$
  with at most $n\xmd(k + 1)^n$ states;
  for every $i$, $0 \leq i < k$, there exists an unambiguous
  subautomaton $\Bc_k^{(i)}$ of $\Bc_k$ which recognises the support
  of $s \monus i$;
  there exists a subautomaton $\Dc_k$ of $\Bc_k$ whose behaviour
  is $s \monus k$.
  \label{t.multi-skim}
\end{theorem}

Here $s \monus k$ is the series obtained from $s$ by subtracting
$k$ to every coefficient larger than $k$ and assigning $0$ to
the others.
In particular, Theorem~\ref{t.multi-skim} says that, if $\Ac$ is
a $k$-ambiguous automaton, then there exists a finite
covering $\Bc_k$ of $\Ac$ and unambiguous subautomata $\Bc_k^{(0)},
\dots, \Bc_k^{(k - 1)}$ of $\Bc_k$ such that the successful
computations of the union $\bigcup_i \Bc_k^{(i)}$ are in bijection
with those of $\Ac$.
Of course, it is not new that $s \monus k$ is a
$\mathbb{N}$-rational series when $s$ is.
This is an old result by Schützenberger which can be proved by iterated
applications of Eilenberg's Cross-Section Theorem~\cite{Eil74}, or of
the construction given in~\cite{Sakarovitch1999}.
But all these methods yield an 
automaton whose size is a tower of exponentials of height $k$.
Theorem~\ref{t.multi-skim} thus answers a problem
left open in~\cite{Sakarovitch1999} with a solution which is better
than the one that was conjectured there.

\begin{figure}[h]
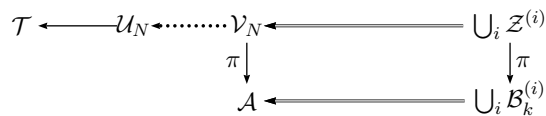

  \centering
  \SmallPicture 
  \VCDraw{%
    \begin{VCPicture}{(0,0.7)(13,-2.5)}
      \ChgStateLineStyle{none}
      \State[\T]{(0,0)}{T}
      \State[\Tpsc]{(3,0)}{Tpsc}
      \State[\Tpsi]{(6,0)}{Tpsi}
      \State[\Aka]{(6,-2)}{Aka}
      \StateVar[\bigcup_i \Asi_k^{(i)}]{(13,-2)}{Asi}
      \StateVar[\bigcup_i \Asil^{(i)}]{(13,0)}{Asil}
      \EdgeR{Tpsc}{T}{}
      \FixEdgeLineDouble{.6}{1}
      \EdgeLineDouble
      \EdgeL{Asi}{Aka}{}
      \EdgeR{Asil}{Tpsi}{}
      \EdgeLineSimple
      \dotted{\EdgeR{Tpsi}{Tpsc}{}}
      \EdgeR{Tpsi}{Aka}{\pi}
      \EdgeL{Asil}{Asi}{\pi}
    \end{VCPicture}
  }
  \caption{Decomposition of a $k$-valued transducer $\T$.
    The simple edge stands for a covering;
    the dotted one represents an input-$k$-ambiguous subautomaton;
    the double ones are immersions;
    $\pi$ is a projection on the underlying input automaton.}
  \label{f.d-dec}
\end{figure}

These coverings together give in two steps a decomposition of a
$k$-valued transducer~$\T$.
First, the lag separation covering of $\T$ yields a
transducer $\Tpsi$ equivalent to $\T$ and whose underlying input
automaton, say $\Ac$, is $k$-ambiguous.
Next, the multi-skimming covering applied to
$\Ac$ yields, as stated in the discussion after
Theorem~\ref{t.multi-skim}, $k$ unambiguous automata $\Bc_k^{(i)}$;
the successful computations of the union of the
$\Bc_k^{(i)}$ are in bijection with those of $\Ac$
and, as the transitions of every $\Bc_k^{(i)}$
map on those of $\Ac$, one can ``lift'' on them
the output of the corresponding transitions of
$\Tpsi$: one thus obtain $k$ unambiguous functional
transducers $\Asil^{(0)}, \dots, \Asil^{(k -  1)}$ decomposing
$\T$ (see Figure~\ref{f.d-dec}).


Our proof goes so to speak in the opposite way that Weber's one:
our first step is to build an input-$k$-ambiguous
transducer from which the decomposition is extracted, whereas the
existence of such a transducer is viewed in~\cite{Weber96} as a
consequence of the decomposition.
Moreover, although both proofs have a very general
idea in common --- a classification of computations from which
at most $k$ successful ones can be distinguished, for every input word
--- the way we do this is completely different.
Indeed, Weber's decomposition is extracted from the
strongly connected components of a graph built on a
preliminary decomposition of $\Tc$ into exponentially many
functional transducers.
We perform a selection among the computations of $\Tc$ according to
a lexicographic ordering on the transitions.

A rough estimation of the complexity (number of states, as a
function on the number of states of $\T$) of this two-step
procedure gives a double exponential: one for the lag separation
covering $\Tpsc$ and other for the multi-skimming covering.
However, a major feature of this construction is that every
computation in the newly built automata corresponds to a computation
in the original transducer --- this is basically what we mean by
structural proof --- which allows to track down the usefulness of
every newly created state.
Then, a careful analysis shows that restricting the constructions to
the $\emph{trim}$ parts of the automata the number of obtained states
is bounded by $2^{\Oc(hLk^4n^{k + 4})}$ states, where $n$ is the
number of states of $\T$, $h$ is the size of the output alphabet and
$L$ is the maximal length of the outputs of the transitions
(Section~\ref{s.dka}).
This is to be compared with the size of Weber's decomposition
described in~\cite{Weber96}, $2^{2^P}$, where $P = p(n + L + h + k)$
is a polynomial whose degree and coefficients do not seem to be easily
derived from the arguments developed there.

The proof of Theorem~\ref{t.gen-weber} starts with the construction of
$k$ unambiguous transducers decomposing the $k$-valued relation $\tau
\theta$.
Next, we show that these transducers induce a decomposition of the set
of successful computations of $\Sc$.
This gives a new set of $k$ finite transducers, not necessarily
unambiguous, which decompose $\Sc$ (Section~\ref{s.gen-dec}).

Finally, let us note that --- as explained in~\cite{Weber96} ---
the improvement in the size of the decomposition from double to
single exponential yields an improvement of the same order for the
complexity of the decision of the equivalence of $k$-valued
transducers.


\section{Preliminaries}
\label{s.prem}

We basically follow the definitions and notation
in~\cite{Bers79,Eil74,Sakarovitch2003}.

The semiring of the nonnegative integers is denoted by $\Nmbb$,
the set of words over a finite alphabet $A$ (the free monoid over $A$)
by $A^*$ and the empty word by $\ew_{A^*}$.
The length of $u \in A^*$  is denoted by $|u|$.
The powerset of a set $X$ is denoted by $\jspart{X}$.

An \emph{automaton} over a monoid $M$ is a labelled directed graph $\Ac = (Q,
M, E, I, T)$ defined by the set $Q$ of vertices, called \emph{states}
and $E$ of edges, called \emph{transitions}, together with two subsets
$I$ and $T$ of $Q$, the initial and final states respectively.
Every transition $e$ in $E$ is associated with a triple $(p, m, q)$ of
$Q \x M \x Q$, specifying its origin, label and end.
Note that we shall explicitly consider cases where \emph{distinct
  transitions} have the \emph{same} origin, label and end, even
though we take the liberty to write $e: p \comput{m} q \in
E$ meaning a transition $e$ associated with $(p, m, q)$.
The automaton $\Ac$ is \emph{finite} if $Q$ and $E$ are finite.

A \emph{computation} in $\Ac$ is a sequence of transitions $c: p_0
\comput{m_1} p_1 \comput{m_2} \dots \comput{m_l} p_l$, also de\-noted as
$p_0 \computaut{m_1  \dots m_l}{\Ac} p_l$.
Its label is $m_1 \dots m_l \in M$ and its length
$l$.
It is \emph{successful} if $p_0 \in I$ and $p_l \in T$.
The \emph{behaviour} of $\Ac$ is the set $\compA \subseteq M$ of labels
of successful computations.
These sets are the family $\Rat M$ of the \emph{rational subsets} of
$M$.

A state of $\Ac$ is \emph{accessible} if it can be reached by
a computation starting at some state of $I$, and \emph{co-accessible}
if some state of $T$ can be reached from it.
The state is \emph{useful} if is both accessible and co-accessible, and
we say that $\Ac$ is \emph{trim} if every state is useful.

If $M$ is a free monoid $A^*$ and the labels of transitions
are letters, then $\Ac$ is a classical automaton over
$A$; we write in this case $\Ac = (Q, A, E, I, T)$.
If $M$ is a product $A^* \times B^*$, then every transition
is labelled by a pair denoted as $u | x$ and
consisting of an input word $u \in A^*$ and an
output one $x \in B^*$; and $\Ac$ is a \emph{transducer} realising a
\emph{rational relation} from $A^*$ to $B^*$.
The image of a word $u \in A^*$ by a transducer is the set of outputs
of successful computations whose input is $u$.
The transducer is called \emph{$k$-valued}, for $k \in \Nmbb$,
if the cardinality of the image of every input word is at
most $k$.

By using classical constructions on automata, every transducer
can be transformed into a \emph{real-time} one: a transducer
whose labels are of form $a | K$, where $a$ is a \emph{letter},
$K \in \Rat B^*$ and $I$ and $T$ are functions from $Q$ to
$\Rat B^*$~\cite{Eil74,Sakarovitch2003}.
For finite image relations we may suppose that the transitions read a
letter and output a single word, and the image of every final
state is $\ew_{B^*}$.
In this case, the transducer is denoted rather as $\T =
(Q, A, B^*, E, I, T)$.

The \emph{underlying input automaton} $\Ac$ of a real-time transducer
$\Tc$ is the (classical) automaton obtained by forgetting the output
of the transitions and replacing the functions $I$ and $T$ by their
domains.
The behaviour of $\Ac$ is the domain of the relation realised by
$\Tc$.

\newsavebox{\SB}
\sbox{\SB}{
 \begin{minipage}[h]{3.8cm}
  \footnotesize
  \begin{displaymath}
    a^n \rightarrow
    \begin{cases}
      1_{B^*} & n = 0 \\
      \{b^n, \, b^{n + 1}\} & n > 0
    \end{cases}
  \end{displaymath}
  \normalsize
\end{minipage}}

\begin{figure}[!h]
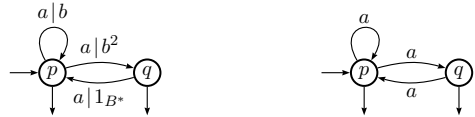

  \centering
  \TinyPicture
  \subfigure{
    \VCDraw{
      \begin{VCPicture}{(-3,1)(6,-.5)}
        \State[p]{(0,0)}{p}
        \State[q]{(3,0)}{q}
        \Initial[w]{p}
        \Final[s]{p}
        \Final[s]{q}
        \ArcL[.5]{p}{q}{\IOL{a}{b^2}}
        \ArcL[.5]{q}{p}{\IOL{a}{1_{B^*}}}
        \LoopN[.5]{p}{\IOL{a}{b}}
      \end{VCPicture}
    }
  }
  \subfigure{
    \VCDraw{
      \begin{VCPicture}{(-3,1)(6,-.5)}
        \RstStateLineStyle
        \State[p]{(0,0)}{pp}
        \State[q]{(3,0)}{qq}
        \Initial[w]{pp}
        \Final[s]{pp}
        \Final[s]{qq}
        \ArcL[.5]{pp}{qq}{a}
        \ArcL[.5]{qq}{pp}{a}
        \LoopN[.5]{pp}{a}
      \end{VCPicture}
    }
  }
  \caption{A $2$-valued real-time transducer $\Tc$ over $\{a\}^* \x
    \{b\}^*$ and its (infinitely ambiguous) underlying input
    automaton.
    The behaviour of $\Tc$ is the relation defined by
    $(1_{A^*})\CompAuto{\Tc} = 1_{B^*}$ and $(a^n)\CompAuto{\Tc} =
    \{b^n, b^{n + 1}\}$ for $n > 0$.}
%
%
  \label{f.T}
\end{figure}


An \emph{$\Nmbb$-automaton} is an automaton labelled by letters with
multiplicities in $\Nmbb$ attached to the transitions and to
initial and final states.
It realises an \emph{$\Nmbb$-rational series}: a function
$s: A^* \rightarrow \Nmbb$ which assigns to $u \in A^*$
a multiplicity given by summing the multiplicites
(product of the multiplicities of transitions) of the
successful computations labelled by $u$.

Every $\Nmbb$-automaton or real-time transducer can be described by
a \emph{matrix representation} $\lmn$, where $\lambda \in S^Q$ ($\nu
\in S^Q$) is a row (column) vector for the multiplicities of the
initial (final) states, $\mu: A^* \rightarrow S^{Q \times Q}$ is a morphism,
$S = \Nmbb$ for $\Nmbb$-automata and $S = \Rat B^*$
for transducers.
The behaviour can be expressed by the function
which maps every $u \in A^*$ to $\lambda \matmul
u\mu \matmul \nu$.
This leads to call \emph{dimension} the set of states of an
automaton.

\begin{figure}[ht]
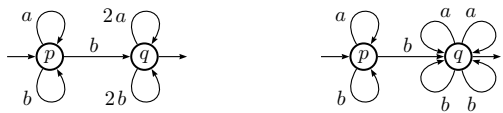

  \centering
  \TinyPicture
  \VCDraw{
    \begin{VCPicture}{(-4,-1.5)(17,1.5)}
      %
      %
      \MediumState
      \State[p]{(0,0)}{p}\State[q]{(3,0)}{q}
      \Initial[w]{p}\Final[e]{q}
      \EdgeL{p}{q}{b}
      \LoopN{p}{a}\LoopN{q}{2\xmd a}
      \LoopS{p}{b}\LoopS{q}{2\xmd b}
      \VCPut{(10,0)}{
        %
        %
        \MediumState
        \State[p]{(0,0)}{pp}\State[q]{(3,0)}{qq}
        \Initial[w]{pp}\Final[e]{qq}
        \EdgeL{pp}{qq}{b}
        \LoopN{pp}{a}
        \LoopS{pp}{b}
        \LoopNW{qq}{a}
        \LoopSW{qq}{b}
        \LoopNE{qq}{a}
        \LoopSE{qq}{b}
      }
    \end{VCPicture}
  }
  \caption{An $\Nmbb$-automaton $\Cc_1$ over $\{a, b\}$,
    on the right-hand side with characteristic transitions
    obtained by splitting multiplicities.
    If $u \in \{a, b\}^*$ is viewed as the writing in the
    binary system of an integer $\overline{u}$ by interpreting $a$ as
    $0$ and $b$ as $1$, then $u \CompAuto{\Cc_1} =
    \overline{u}$.}
  \label{f.A}
\end{figure}

It will be useful 
to consider $\Nmbb$-automata whose transitions
are \emph{characteristic}, that is, with multiplicity $1$.
Every $\Nmbb$-automaton can be
transformed into such a one by splitting every transition
with multiplicity $l > 0$ into a set of $l$ 
characteristic ones (Figure~\ref{f.A}).

A \emph{morphism} from $\Bc = (R, M, F, J, U)$ to $\Ac = (Q,
M, E, I, T)$, denoted
by $\varphi: \Bc \rightarrow \Ac$, is a pair of mappings $R
\rightarrow Q$ and $F \rightarrow E$, both denoted by $\varphi$, such
that $J\varphi \subseteq I$, $U\varphi \subseteq T$ and for every $e \in
F$, if $e$ is associated with $(p, m, q)$, then $e\varphi$ is
associated with $(p\varphi, m, q\varphi)$.
We say that $\varphi$ is a \emph{covering} if $\varphi$ induces a bijection
between the outgoing transitions of $p$ and $p\varphi$,
$I$ is in bijection with $J$ and $T\varphi^{-1} = U$.
An \emph{immersion} is by definition a subautomaton of a covering.
These conditions imply that every
successful computation of $\Bc$ maps to a successful
computation of $\Ac$, and thus $\CompAuto{\Bc} \subseteq
\CompAuto{\Ac}$.
In the case of coverings, there is indeed a
bijection between the successful computations and thus $\CompAuto{\Bc}
= \CompAuto{\Ac}$~\cite{sakarovitch98}.

A covering of the split form of an $\Nmbb$-automaton $\Ac$ is the split
form of an \emph{$\Nmbb$-covering} of $\Ac$,
see~\cite{BLS06,Sakarovitch2003,Sakarovitch1999} for the definition of
the latter.
The $\Nmbb$-series realised by an $\Nmbb$-automata and any
of its $\Nmbb$-coverings are the same.


\section{Lexicographic coverings}
\label{s.lex-ord}

The idea of the two coverings we are going to define
is to order lexicographically computations of automata, inasmuch as it
can be done with words on some alphabet.
Here, the alphabet is the set of transitions, and
computations are seen as words on it.


\subsection{The lexicographic ordering of computations}

Let $\Ac = (Q, A, E, I, T)$ be a classical automaton.
Fix a (partial) ordering $\llex$ on $E$ such that transitions are
comparable iff they have the same label and origin.
This ordering is extended on $E^*$ and thus on the computations of
$\Ac$ in such a way that it can be called a \emph{lexicographic ordering
of the computations}:
$c = e_1 e_2 \dots e_l e_{l + 1} \dots e_n$
and $d = e'_1 e'_2 \dots e'_l e'_{l + 1} \dots e'_m$ ($e_i, e'_j
\in E$ for $1 \leq i \leq n$ and $1 \leq j \leq m$) are such that
$c \llex d$ iff $c$ and $d$ have the same label (thus $m = n$)
and there exists $l$ such that
 $e_i = e'_i$ for $1 \leq i \leq l - 1$ and $e_l \llex
e'_l$.

\begin{figure}[h]
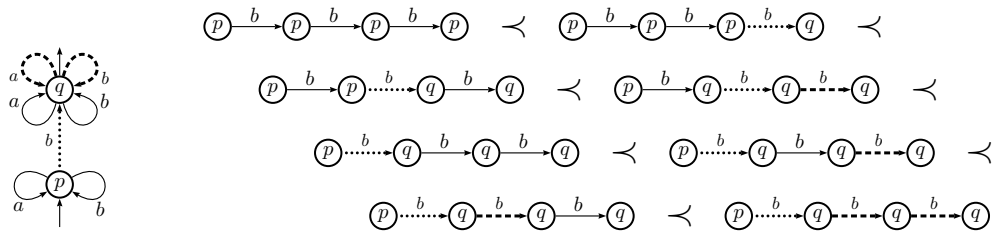

  \centering
  \TinyPicture 
  \VCDraw{%
    \begin{VCPicture}{(-6,-2)(26,5)}
      \newlength{\rsTL}\setlength{\rsTL}{2.5cm}
      %
      %
      %
      \MediumState
      \VCPut{(-4,0)}{%
      \State[p]{(0,-0.5)}{pp}\State[q]{(0,2.5)}{qq}
      }%
      \Initial[s]{pp}\Final[n]{qq}
      \dotted{\EdgeL{pp}{qq}{b}}
      \LoopW[.75]{pp}{a}
      \LoopE[.75]{pp}{b}
      \LoopSW[.75]{qq}{a}
      \LoopSE[.75]{qq}{b}
      \dashed{\LoopNW[.75]{qq}{a}}
      \dashed{\LoopNE[.75]{qq}{b}}
      \RstEdgeLineStyle
      %
      %
      \VCPut{(1,4.5)}{%
      \VCPut{(0,0)}{%
         \VCPut{(0,0)}{%
            \State[p]{(0\rsTL,0)}{P1}\State[p]{(1\rsTL,0)}{P2}%
            \State[p]{(2\rsTL,0)}{P3}\State[p]{(3\rsTL,0)}{P4}}%
               \ChgStateLabelScale{1.8}%
               \VCPutStateLabel{(3.75\rsTL,0)}{\llex}%
               \VCPutStateLabel{(8.25\rsTL,0)}{\llex}%
               \RstStateLabelScale%
         \VCPut{(4.5\rsTL,0)}{%
            \State[p]{(0\rsTL,0)}{P1A}\State[p]{(1\rsTL,0)}{P2A}%
            \State[p]{(2\rsTL,0)}{P3A}\State[q]{(3\rsTL,0)}{Q4A}}%
       }%
      \VCPut{(0.7\rsTL,-2)}{%
         \VCPut{(0,0)}{%
            \State[p]{(0\rsTL,0)}{P1B}\State[p]{(1\rsTL,0)}{P2B}%
            \State[q]{(2\rsTL,0)}{Q3B}\State[q]{(3\rsTL,0)}{Q4B}}%
               \ChgStateLabelScale{1.8}%
               \VCPutStateLabel{(3.75\rsTL,0)}{\llex}%
               \VCPutStateLabel{(8.25\rsTL,0)}{\llex}%
               \RstStateLabelScale%
         \VCPut{(4.5\rsTL,0)}{%
            \State[p]{(0\rsTL,0)}{P1C}\State[q]{(1\rsTL,0)}{Q2C}%
            \State[q]{(2\rsTL,0)}{Q3C}\State[q]{(3\rsTL,0)}{Q4C}}%
      }%
      \VCPut{(1.4\rsTL,-4)}{%
         \VCPut{(0,0)}{%
            \State[p]{(0\rsTL,0)}{P1D}\State[q]{(1\rsTL,0)}{Q2D}%
            \State[q]{(2\rsTL,0)}{Q3D}\State[q]{(3\rsTL,0)}{Q4D}}%
               \ChgStateLabelScale{1.8}%
               \VCPutStateLabel{(3.75\rsTL,0)}{\llex}%
               \VCPutStateLabel{(8.25\rsTL,0)}{\llex}%
               \RstStateLabelScale%
         \VCPut{(4.5\rsTL,0)}{%
            \State[p]{(0\rsTL,0)}{P1E}\State[q]{(1\rsTL,0)}{Q2E}%
            \State[q]{(2\rsTL,0)}{Q3E}\State[q]{(3\rsTL,0)}{Q4E}}%
      }%
      \VCPut{(2.1\rsTL,-6)}{%
         \VCPut{(0,0)}{%
            \State[p]{(0\rsTL,0)}{P1F}\State[q]{(1\rsTL,0)}{Q2F}%
            \State[q]{(2\rsTL,0)}{Q3F}\State[q]{(3\rsTL,0)}{Q4F}}%
               \ChgStateLabelScale{1.8}%
               \VCPutStateLabel{(3.75\rsTL,0)}{\llex}%
               \RstStateLabelScale%
         \VCPut{(4.5\rsTL,0)}{%
            \State[p]{(0\rsTL,0)}{P1G}\State[q]{(1\rsTL,0)}{Q2G}%
            \State[q]{(2\rsTL,0)}{Q3G}\State[q]{(3\rsTL,0)}{Q4G}}%
      }%
      }%
      %
      \EdgeL{P1}{P2}{b}
      \EdgeL{P2}{P3}{b}
      \EdgeL{P3}{P4}{b}
      \EdgeL{P1A}{P2A}{b}
      \EdgeL{P2A}{P3A}{b}
      \dotted{\EdgeL{P3A}{Q4A}{b}}
      \EdgeL{P1B}{P2B}{b}
      \dotted{\EdgeL{P2B}{Q3B}{b}}
      \EdgeL{Q3B}{Q4B}{b}
      \EdgeL{P1C}{Q2C}{b}
      \dotted{\EdgeL{Q2C}{Q3C}{b}}
      \dashed{\EdgeL{Q3C}{Q4C}{b}}
      \dotted{\EdgeL{P1D}{Q2D}{b}}
      \EdgeL{Q2D}{Q3D}{b}
      \EdgeL{Q3D}{Q4D}{b}
      \dotted{\EdgeL{P1E}{Q2E}{b}}
      \EdgeL{Q2E}{Q3E}{b}
      \dashed{\EdgeL{Q3E}{Q4E}{b}}
      \dotted{\EdgeL{P1F}{Q2F}{b}}
      \dashed{\EdgeL{Q2F}{Q3F}{b}}
      \EdgeL{Q3F}{Q4F}{b}
      \dotted{\EdgeL{P1G}{Q2G}{b}}
      \dashed{\EdgeL{Q2G}{Q3G}{b}}
      \dashed{\EdgeL{Q3G}{Q4G}{b}}
      \RstStateLineStyle
    \end{VCPicture}
  }
  \caption{A lexicographic ordering between the computations of
    $\Cc_1$ labelled by $bbb$ and starting at $p$.
    Solid transitions are smaller than the dotted and dashed
    ones.}
  \label{f.loc}
\end{figure}

The definitions for other kinds of automata are similar but, in order
to give them the wanted meaning, a little bit more delicate: for
$\Nmbb$-automata, the ordering is put on the split form, and for
real-time transducers, on the underlying input automaton.\footnote{In
  order to ease the explanation, we shall describe the constructions
  for automata with a single initial state. Computations starting at
  distinct initial states become ordered by extending $\llex$ to new
  transitions $i \comput{1} p$ starting at a
  ``hidden'' initial state $i$, for every $p \in I$.
  Initial multiplicities can be treated similarly.}


\subsection{The multi-skimming covering of an $\Nmbb$-automaton}
\label{s.lex-cov}

The aim of the multi-skimming covering of an
$\Nmbb$-automaton $\Ac = (Q, A^*, E, i, T)$  is to count, for
every successful computation, the number of the smaller ones according
to $\llex$.

Let $\smtrans: E \rightarrow \Nmbb^Q$ be the function from transitions
to $\Nmbb$-vectors indexed by $Q$ defined by
$(e \smtrans)_r = \card { \{ f \in E \mid f: p
      \comput{a} r \mbox{ and } f \llex e \}}$, for
$e: p \comput{a} q \in E$ and $r \in Q$.

\begin{definition}
  The multi-skimming covering of $\Ac$
  is the (infinite) $\Nmbb$-automaton $\Bc$ of dimension $Q \times
  \Nmbb^Q$ defined as follows:
  \begin{itemize}

  \item the initial state is $(i, \, \zv)$ (where
    $\zv$ is the zero vector);

  \item the final states are $T \times \Nmbb^Q$;

  \item for every $(p, \, \Nvec{v}) \in Q
    \times \Nmbb^Q$ and every $e: p \comput{a} q \in E$,
    $(p, \Nvec{v}) \comput{a} (q, \Nvec{v} \matmul a\mu
      + e \smtrans)$  is a transition of $\Bc$
    (where $\mu$ is the morphism of the matrix representation of
    $\Ac$). \varqed

  \end{itemize}

  \label{d.lex-cov}
\end{definition}

It follows from this definition that
for every state $(p, \Nvec{v})$ of $\Bc$, the outgoing
transitions of $(p, \Nvec{v})$ are in bijection with those of $p$.
Thus, the projection $\varphi$ of $\Bc$ on the first component is an
$\Nmbb$-covering of $\Ac$.
The property below follows by induction on the length of
computations\footnote{Computations of coverings will be represented
  with capital letters.}:

\begin{property}
  Let $C: (i, \zv) \computaut{u}{\Bc} (p, \Nvec{v})$
  be a computation.
  For every $q \in Q$, $\Nvec{v}_q$ is the number of computations
  $d: i \computaut{u}{\Ac} q$ such that $d \llex C\varphi$ (where
  $C\varphi$ is the projection of $C$ on $\Ac$). \qed
  \label{py.lex-cov}
\end{property}

\begin{figure}[h]
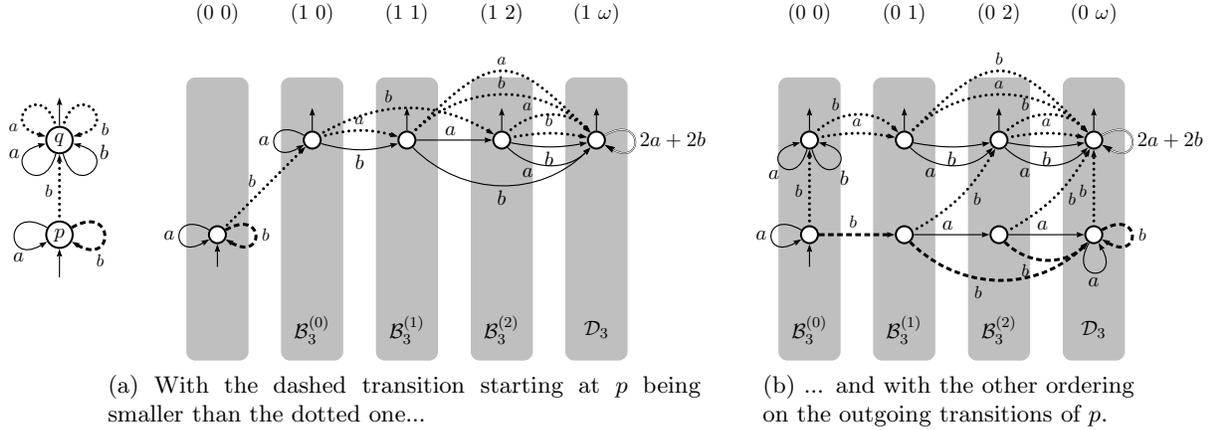

  \centering
  \TinyPicture 
  \subfigure[With the dashed transition starting at $p$ being smaller than the
  dotted one...]{
    \VCDraw{%
      \begin{VCPicture}{(1,-3.5)(21,6)} 
        %
        %
        \rput{0}(4,3){\psframe[linewidth=0pt,linecolor=white,fillstyle=solid,fillcolor=lightgray,framearc=.3](0,2)(2,-7)}
        \rput{0}(7,3){\psframe[linewidth=0pt,linecolor=white,fillstyle=solid,fillcolor=lightgray,framearc=.3](0,2)(2,-7)}
        \rput{0}(10,3){\psframe[linewidth=0pt,linecolor=white,fillstyle=solid,fillcolor=lightgray,framearc=.3](0,2)(2,-7)}
        \rput{0}(13,3){\psframe[linewidth=0pt,linecolor=white,fillstyle=solid,fillcolor=lightgray,framearc=.3](0,2)(2,-7)}
        \rput{0}(16,3){\psframe[linewidth=0pt,linecolor=white,fillstyle=solid,fillcolor=lightgray,framearc=.3](0,2)(2,-7)}
        \ChgStateLineStyle{none}
        \ChgStateFillColor{lightgray}
        \State[\Bc_3^{(0)}]{(8,-3)}{B0}
        \State[\Bc_3^{(1)}]{(11,-3)}{B1}
        \State[\Bc_3^{(2)}]{(14,-3)}{B2}
        \State[\Dc_3]{(17,-3)}{C}
        \ChgStateLineStyle{solid}
        \ChgStateFillColor{white}
        %
        %
        \MediumState
        \State[p]{(0,0)}{pp}\State[q]{(0,3)}{qq}
        \Initial[s]{pp}\Final[n]{qq}
        \dotted{\EdgeL{pp}{qq}{b}}
        \LoopW[.75]{pp}{a}
        \dashed{\LoopE[.75]{pp}{b}}
        \LoopSW[.75]{qq}{a}
        \LoopSE[.75]{qq}{b}
        \dotted{\LoopNW[.75]{qq}{a}}
        \dotted{\LoopNE[.75]{qq}{b}}
        %
        %
        %
        \ChgStateLineStyle{none}
        \State[(0 \,\, 0)]{(5,7)}{v00}
        \State[(1 \,\, 0)]{(8,7)}{v10}
        \State[(1 \,\, 1)]{(11,7)}{v11}
        \State[(1 \,\, 2)]{(14,7)}{v12}
        \State[(1 \,\, \omega)]{(17,7)}{v1w}
        \SmallState
        \ChgStateLineStyle{solid}
        \State{(5,0)}{p00}
        \State{(8,3)}{q10}
        \State{(11,3)}{q11}
        \State{(14,3)}{q12}
        \State{(17,3)}{q1w}
        \ChgEdgeLineStyle{dotted}
        \dotted{\EdgeL{p00}{q10}{b}}
        \dotted{\ArcL[.5]{q10}{q11}{a}}
        \dotted{\LArcL{q10}{q12}{b}}
        \SetArcAngle{10}
        \dotted{\ArcL[.5]{q12}{q1w}{b}}
        \SetArcAngle{45}
        \dotted{\ArcL[.3]{q12}{q1w}{a}}
        \SetArcAngle{45}
        \dotted{\ArcL[.5]{q11}{q1w}{b}}
        \SetArcCurvature{1.3}
        \dotted{\ArcL[.5]{q11}{q1w}{a}}
        \SetArcCurvature{0.8}
        \ArcR[.5]{q11}{q1w}{b}
        \SetArcAngle{15}
        \SetArcCurvature{0.8}
        \ArcR[.5]{q10}{q11}{b}
        \EdgeL{q11}{q12}{a}
        \SetArcAngle{10}
        \ArcR[.5]{q12}{q1w}{b}
        \SetArcAngle{45}
        \ArcR[.3]{q12}{q1w}{a}
        \LoopW[.5]{q10}{a}
        \LoopW[.5]{p00}{a}
        \dashed{\LoopE[.5]{p00}{b}}
        \Initial[s]{p00}
        \Final[n]{q10}
        \Final[n]{q11}
        \Final[n]{q12}
        \Final[n]{q1w}
        \EdgeLineDouble \LoopE[.5]{q1w}{2a + 2b}
      \end{VCPicture}
    }
  } 
%
%
  \subfigure[... and with the other ordering on the outgoing
  transitions of $p$.]{
    \VCDraw{%
      \begin{VCPicture}{(3,-3.5)(16,6)} 
        %
        %
        \rput{0}(4,3){\psframe[linewidth=0pt,linecolor=white,fillstyle=solid,fillcolor=lightgray,framearc=.3](0,2)(2,-7)}
        \rput{0}(7,3){\psframe[linewidth=0pt,linecolor=white,fillstyle=solid,fillcolor=lightgray,framearc=.3](0,2)(2,-7)}
        \rput{0}(10,3){\psframe[linewidth=0pt,linecolor=white,fillstyle=solid,fillcolor=lightgray,framearc=.3](0,2)(2,-7)}
        \rput{0}(13,3){\psframe[linewidth=0pt,linecolor=white,fillstyle=solid,fillcolor=lightgray,framearc=.3](0,2)(2,-7)}
        \ChgStateLineStyle{none}
        \ChgStateFillColor{lightgray}
        \State[\Bc_3^{(0)}]{(5,-3)}{B0}
        \State[\Bc_3^{(1)}]{(8,-3)}{B1}
        \State[\Bc_3^{(2)}]{(11,-3)}{B2}
        \State[\Dc_3]{(14,-3)}{C}
        \ChgStateLineStyle{solid}
        \ChgStateFillColor{white}
        %
        %
        \ChgStateLineStyle{none}
        \State[(0 \,\, 0)]{(5,7)}{v00}
        \State[(0 \,\, 1)]{(8,7)}{v01}
        \State[(0 \,\, 2)]{(11,7)}{v02}
        \State[(0 \,\, \omega)]{(14,7)}{v0w}
        \SmallState
        \ChgStateLineStyle{solid}
        \ChgEdgeLineStyle{solid}
        \State{(5,0)}{p00}
        \State{(8,0)}{p01}
        \State{(11,0)}{p02}
        \State{(14,0)}{p0w}
        \State{(5,3)}{q00}
        \State{(8,3)}{q01}
        \State{(11,3)}{q02}
        \State{(14,3)}{q0w}
        \LoopW[.5]{p00}{a}
        \dotted{\EdgeL{p00}{q00}{b}}
        \LoopSW[.5]{q00}{a}
        \LoopSE[.5]{q00}{b}
        \EdgeL{p01}{p02}{a}
        \dotted{\ArcR[.6]{p01}{q02}{b}}
        \EdgeL{p02}{p0w}{a}
        \dotted{\ArcR[.6]{p02}{q0w}{b}}
        \SetArcAngle{10}
        \ArcR[.5]{q02}{q0w}{b}
        \ArcR[.5]{q01}{q02}{b}
        \SetArcAngle{45}
        \ArcR[.3]{q02}{q0w}{a}
        \ArcR[.3]{q01}{q02}{a}
        \LoopS[.5]{p0w}{a}
        \dotted{\EdgeL{p0w}{q0w}{b}}
        \dashed{\EdgeL{p00}{p01}{b}}
        \SetArcAngle{10}
        \dotted{\ArcL[.5]{q00}{q01}{a}}
        \dotted{\ArcL[.5]{q02}{q0w}{a}}
        \SetArcAngle{45}
        \dotted{\ArcL[.3]{q00}{q01}{b}}
        \dotted{\ArcL[.3]{q02}{q0w}{b}}
        \dotted{\ArcL[.5]{q01}{q0w}{a}}
        \SetArcCurvature{1.3}
        \dotted{\ArcL[.5]{q01}{q0w}{b}}
        \SetArcCurvature{0.8}
        \dashed{\ArcR{p01}{p0w}{b}}
        \dashed{\ArcR[.3]{p02}{p0w}{b}}
        \SetArcAngle{15}
        \dashed{\LoopE[.5]{p0w}{b}}
        \Initial[s]{p00}
        \Final[n]{q00}
        \Final[n]{q01}
        \Final[n]{q02}
        \Final[n]{q0w}
        \EdgeLineDouble \LoopE[.5]{q0w}{2a + 2b}
      \end{VCPicture}
    }
  }
  \vspace*{-0.3cm}
  
  \caption{The multi-skimming covering at layer $k = 3$ for $\Cc_1$
    with two different orderings of the transitions starting at $p$.
    States of the covering are pairs $(r, \Nvec{v})$, where $r$ is a
    state of the automaton (horizontal projection) and $\Nvec{v}$ is an
    $\Nmbb_3$-vector indexed by $\{p, \, q\}$ in that order (vertical
    projection).
    Solid transitions leaving $q$ are smaller than the dotted ones in
    both coverings.
    The double transitions stand for four transitions.
    The automata $\Bc_3^{(i)}$ recognising the support of $ s \monus
    i$ and $\Dc_3$ recognising $s \monus 3$ are given by keeping as
    final exactly one final state at the indicated columns.}
  \label{f.multi-skimming}
\end{figure}

We define as above the (finite) automaton $\Bc_k$ satisfying
Theorem~\ref{t.multi-skim}; the difference is that
it counts \emph{until $k - 1$}.
Let $\Nmbb_k = \{0, \, \dots, \, k - 1, \, \omega\}$ be the quotient
semiring of $\Nmbb$ given by the relation $k =  k + 1$
($\omega$ is the class of $k$ and plays the role of an
infinity).
The dimension of $\Bc_k$ is $Q \times \Nmbb_k^Q$; transitions and
initial and final states are defined as in Definition~\ref{d.lex-cov},
but the matrix operations
$\Nvec{v} \matmul a\mu + e \smtrans$ are made in $\Nmbb_k$.
The morphism $\Nmbb \rightarrow \Nmbb_k$ induces an
$\Nmbb$-quotient $\Bc \rightarrow \Bc_k$, and as noted,
$\Bc_k$ is an $\Nmbb$-covering of $\Ac$.
Figure~\ref{f.multi-skimming} shows an example.

By induction on the length of computations, we have:

\begin{property}
  Let $C: (i, \zv) \computaut{u}{\Bc_k} (p, \Nvec{v})$
  be a computation.
  For every $q \in Q$, $\Nvec{v}_q$ is the number of computations
  $d: i \computaut{u}{\Ac} q$ such that $d \llex C\varphi$,
  if this number is smaller than $k$, or it is $\omega$
  otherwise. \qed
  \label{py.lex-cov-quot}
\end{property}

\proof[Proof of Theorem~\ref{t.multi-skim}]

In view of Property~\ref{py.lex-cov-quot}, we can obtain the subautomata
$\Bc_k^{(i)}$ of $\Bc_k$ by erasing the
condition of being final of some final states of $\Bc_k$:
each $\Bc_k^{(i)}$ is defined by choosing as final
only the states $(p, \Nvec{v}) \in T \times \Nmbb_k^Q$ such that
$\sum_{q \in T} \Nvec{v}_q = i$; $\Dc_k$ is the subautomaton of
$\Bc_k$ defining as final the states $(p, \Nvec{v}) \in T \times
\Nmbb_k^Q$ such that $\sum_{q \in T} \Nvec{v}_q = \omega$. \qed


\subsection{The lag separation covering of a real-time transducer}
\label{s.N-sep}

Let $\T = (Q, A, B^*, E, i, T)$ be a real-time
transducer.
We aim with the lag separation covering of $\T$ at
a selection between computations of this transducer with
\emph{same input and same output} (stated in Property~\ref{py.psi}).
This will be useful in Section~\ref{s.dec} to construct a
input-$k$-ambiguous transducer from a $k$-valued one.

It is not possible in general to build a finite expansion which allows
to select exactly one
computation for each pair of words in the relation realised by the
transducer, for this would lead to an unambiguous 
transducer and there exist rational relations
which are inherently ambiguous.
The idea is to fix a parameter~$N$ and compare only
computations such that \emph{the differences of lengths of outputs
  along them (their ``lag'') are bounded by $N$}.

At first, let us recall the \emph{Lead or Delay action},
defined in~\cite{BCPS03} to describe differences of words.
We restate it in a slightly different form, based on
the \emph{free group} $\FG{B}$ generated by $B$:
the quotient of $(B \cup \Bbar)^*$ by the relations
$x\xmd\xbar = \xbar\xmd x = \ew_{B^*}$ ($x \in B$), where
$\Bbar$ a disjoint copy of $B$.
The inverse of $u \in B^*$, denoted
by $\ubar$, is the mirror image of $u$ with barred letters.
We denote by $\ewfg$ the empty word of $\FG{B}$ (which is the class of
the empty word of $B$).
Let $\LDset{B} = B^* \cup \Bbar^* \cup \{\ldzero\}$,
where $\ldzero$ is a new element, a zero, not in $\FG{B}$,
and $\inD: \FG{B} \cup \{\ldzero\} \rightarrow \LDset{B}$ be the
function $w\inD = w$, if $w \in \LDset{B}$, and $w \inD = \ldzero$
otherwise.

\begin{definition}
  The Lead or Delay Action of $\BeBe$ on $\LDset{B}$ is defined
  by $\msp w \lda (x, y) = (\xbar\xmd w\xmd y)\inD \msp$, 
  $w \in \LDset{B}$, $(x, y) \in \BeBe$ (the product is taken
  with the rules $\msp\ldzero\xmd x = x\xmd \ldzero
  =\ldzero$). \varqed
  \label{d.lda}
\end{definition}

Intuitively, $\msp\ewfg \lda (x, y)\msp$ represents the ``difference''
of the words $x$ and $y$, being a positive word if $x$ is a prefix of
$y$ (the \emph{lead} of $y$ with respect to $x$), a negative word if
$y$ is a prefix of $x$ (the \emph{delay} of $y$ with respect to $x$),
and $\ldzero$ if $x$ and $y$ are not prefixes of a common word.

\begin{definition}
  Let $c: p \comput{u \mid x} q$ and $d: p' \comput{u \mid
    y} q'$ be two computations of $\T$ with the same input $u$.
  As $\T$ is a real-time transducer, $c$ and $d$ have the same
  length.
  We define their \emph{Lead or Delay}, denoted by $\LD{c}{d}$, as the element
    $\ewfg \lda (x, y)$ of $\LDset{B}$, and
    if $\LD{c}{d} \neq \ldzero$, their \emph{lag} as the integer
      $\lag{c}{d} = \max \{|\LD{c'}{d'}| \mid c', d' \mbox{ prefixes of }
      c, d \mbox{ with the same length}\}$. \varqed
  \label{d.ld-lag}
\end{definition}

Similarly to the multi-skimming covering, the states of
the lag separation covering of $\T$ carry vectors
indexed by $Q$.
But in this case the ``stored information'' is the Lead or Delay
between any computation and those which are smaller.
Let $\smtrans: E \rightarrow
\jspart{\LDset{B}}^Q$ be the function given by
$  (e \smtrans)_r = \{ (\xbar y) \inD \, \mid \, f: p
    \comput{a | y} r \in E, \, f \llex e \}$, for
$e: p \comput{a | x} q \in E$ and $r \in Q$.

\begin{definition}
  The lag separation covering of $\T$ is the (infinite) real-time
  transducer $\Uc = (R, A, B^*, F, j, U)$ defined by
  \begin{itemize}

  \item $R = Q \x \jspart{\LDset{B}}^Q$;

  \item $j = (i, \zv)$ (where $\zv$ is the vector whose
    entries are all equal to~$\emptyset$);

  \item $U = T \x R$;

  \item for every $(p, \WSvec{v}) \in R$ and every $e: p\comput{a |
      x} q \in E$, $(p, \WSvec{v}) \comput{a | x} 
    \Tran{q, (\xbar \cdot \WSvec{v} \lda a\mu + e \smtrans) \inD}$ is
    a transition in $F$ (where $\mu$ is the morphism of the matrix
    representation of $\T$, $\xbar \cdot \WSvec{v}$ is the vector
    obtained by multiplying on the left every entry of $\WSvec{v}$ by
    $\xbar$, and $\inD$ is extended componentwise to vectors in
    $\jspart{\LDset{B}}^Q$).
    \varqed
  \end{itemize}
  \label{d.sep-cov}
\end{definition}

As before, for every state $(p, \WSvec{v})$ of $\Uc$, there is a
bijection between the outgoing transitions of $(p, \WSvec{v})$ and
those of $p$: the projection $\varphi$ of $\Uc$ on the first
component is a covering on $\T$.
By induction on the length of computations, we have:

\begin{property}
  Let $C: (i, \zv) \comput{u \mid x} (p,
  \WSvec{v})$ be a computation of $\Uc$.
  For every state $q$ of $\T$,  $\WSvec{v}_q$ is
  the set of Lead or Delay of $C\varphi$ (the projection of $C$
  on $\T$) and any computation of $\T$ smaller than $C\varphi$ and
  which ends in $q$: $\WSvec{v}_q = \{
    \LD{C\varphi}{d} \mid d: i \computaut{u \mid y}{\T} q, \,\, d
    \llex C\varphi \}$. \qed
  \label{py.pre_lsc}
\end{property}

In order to build the announced selection of computations of $\T$,
we define a ``bounded'' lag separation covering where
only the computations with lag bounded by $N$ are compared,
so that only words in $\lagN = B^{\leq N} \cup \Bbar^{\leq N}$ are
``stored'' in the entries of the vectors $\WSvec{v}$.
Let $\inD_N: \FG{B} \rightarrow \lagN \cup \{\ldzero\}$ be the function
defined by $w\inD_N = w$, if $w \in \lagN$, and $w \inD_N = \ldzero$
otherwise.
The element $\ldzero$ is intentionally omitted from $\lagN$ in order to
simplify the writing of Property~\ref{py.lsc}, and in
the extension of $\inD_N$ to $\jspart{\FG{B}}^Q$ the
image of a word not in $\lagN$ will be seen as the empty set so
that for $\WSvec{v} \in
\jspart{\FG{B}}^Q$, $\WSvec{v} \inD_N$ is a vector in
$\jspart{\lagN}^Q$ (which does not contain $\ldzero$ in any of its
entries).
We define $\Tpsc$ as the (finite) transducer constructed as in
Definition~\ref{d.sep-cov}, but with states and transitions given by:
\begin{displaymath}
  R = Q \x \jspart{\lagN}^Q, \qquad \quad
  \forall \, (p, \WSvec{v}) \in R, \,
  \forall \, e: p\comput{a |
    x} q \in E \quad (p, \WSvec{v}) \comput{a | x}
  \Tran{q, (\xbar \cdot \WSvec{v} \cdot a\mu + e \smtrans) \inD_N}
  \in  F.
\end{displaymath}
Due to the fact that $\inD_N$ is not a morphism, it is not true
in general that $\Uc$ is a covering of $\Tpsc$; but $\Tpsc$ is
another covering of $\T$.
By induction we have (see
Figure~\ref{f.dec-lsc}):

\begin{property}
  Let $C: (i, \zv) \comput{u \mid x} (p,
  \WSvec{v})$  be a computation of $\Tpsc$.
  For every state $q$ of $\T$, $\WSvec{v}_q = \{
    \LD{C\varphi}{d} \mid d: i \computaut{u \mid y}{\T} q, \,\,\,
    \mbox{$x,y$ prefixes of a common word}, \,\,
    d \llex C\varphi, \,\,
    \lag{C\varphi}{d} \leq N \}$. \qed
  \label{py.lsc}
\end{property}

The wanted selection is a consequence of Property~\ref{py.lsc} and
can be stated as follows:

\begin{property}
  Let $\Tpsi$ be the subtransducer of $\Tpsc$
  obtained by removing the property of being final of every state
  $(p,\WSvec{v}) \in T \times R$ such that $\ewfg \in
  \mathsf{v}_t$ for some $t \in T$. 
  A computation $C$ of $\Tpsi$ is successful if, and only if,
  $C\varphi$ is successful in $\T$ and for every successful
  computation $d$ of $\T$ smaller than $C\varphi$ with (same input
  and) same output, $\lag{C\varphi}{d} > N$. \qed
  \label{py.psi}
\end{property}

\emph{The transducers $\T$ and $\Tpsi$ are equivalent}: if
$(u, x)$ is in the behaviour of $\T$,
the smallest successful computation of $\T$ labelled by $(u, x)$ is
the projection of a successful one in $\Tpsi$.

The following remark on the trim part of $\Tpsi$ will be useful for
the evaluation of the size of the decomposition
(Section~\ref{s.dka}).

\begin{property}
  Let $\T$ be a trim and $k$-valued transducer with $n$ states,
  and whose output alphabet has $h$ letters.
  The number of useful states of $\Tpsi$ is bounded by $2^{2hNk^2n}$.
  \label{py.size-dec-1}
\end{property}
\proof
We write $\jsPartK{X}{l}$ for
the set of the subsets with at most $l$ elements of a set $X$.
Clearly, $\card{\jsPartK{X}{l}} \leq \card{X}^{l^2}$.
The hypothesis that $\T$ is trim and $k$-valued together with
Property~\ref{py.lsc} imply that the vectors in the useful states of
$\Tpsi$ have in every coordinate at most $k$ words, thus
these states belong to $Q \times
\jsPartK{\lagN}{k}^Q$.
The cardinality of this set is at most $n
\cdot \left ( \card{\lagN}^{k^2} \right )^n \leq n \cdot
\left ( (2h)^{Nk^2} \right)^n$.
This is clearly bounded by $2^{2hNk^2n}$. \qed

\section{Decomposing a $k$-valued rational relation}
\label{s.dec}

As said in the introduction, we first prove a result
for $k$-valued transducers:

\begin{theorem}
  Any $k$-valued transducer is equivalent to an input-$k$-ambiguous
  one.
  \label{t.k-amb}
\end{theorem}

This will be established by the lag separation covering: for some
adequate $N$, $\Tpsi$ is input-$k$-ambiguous
(Proposition~\ref{p.N-sep-k-amb}).
Next, Theorem~\ref{t.weber} is proved by applying
the multi-skimming covering on
the underlying input automaton of $\Tpsi$ (Section~\ref{s.dka}).


\subsection{From a $k$-valued transducer to an input-$k$-ambiguous one}
\label{s.kvka}

\begin{proposition}
  Let~$\Tc$ be a real-time transducer with $n$ states and lengths of outputs
  of transitions bounded by $L$.
  If $\T$ is $k$-valued, then for $N \geq L\xmd n^{k + 1}$
  $\Tpsi$ is input-$k$-ambiguous.
  \label{p.N-sep-k-amb}
\end{proposition}

The crux of the proof is a combinatorial property stated in Theorem~2.2
of~\cite{Weber96}, and restated here as Lemma~\ref{l.cycle-lda}.
In this lemma, $\Tck$ is the cartesian product of $\T$ by
itself $k + 1$ times, a natural generalisation of the
squaring of $\T$ defined in~\cite{BCPS03} to establish the
decidability of the functionality of transducers.
In $\T^2$, every computation corresponds to a pair of computations of $\T$
\emph{with the same input}; in $\Tck$, every computation corresponds
then to a $(k + 1)$ tuple of computations of $\T$ \emph{with the same
  input} (this construction is heavily used in~\cite{SakaSouzaLATIN08}
to give a new proof of the decidability of $k$-valuedness).

\begin{lemma}[Weber~\cite{Weber96}]
  If $\T$ is $k$-valued, then for every successful computation
  $\computvec{c}$ of $\Tck$ there
  exists a pair $i, j$ of coordinates such that the projections
  $\computvec{c}_i$ and $\computvec{c}_j$ satisfy
  $\LD{\computvec{c}_i}{\computvec{c}_j} = \ew$ (that is,
  $\computvec{c}_i$ and $\computvec{c}_j$ have the same output) and
  $\lag{\computvec{c}_i}{\computvec{c}_j} < Ln^{k + 1}$. \qed
  \label{l.cycle-lda}
\end{lemma}

A concise proof for Lemma~\ref{l.cycle-lda} can be derived from
a property of the Lead or Delay action stated in Lemma 5
of~\cite{BCPS03}.
Although not so long, it is omitted due to space constraints.

\proof[Proof of Proposition~\ref{p.N-sep-k-amb}]
Fix $N \geq Ln^{k + 1}$.
By Property~\ref{py.psi}, distinct successful computations of $\Tpsi$
with the same input either output distinct words or have
a lag greater than $Ln^{k + 1}$.
Hence $k + 1$ distinct successful computations of $\Tpsi$ with the
same input word would project on a $(k + 1)$-tuple of computations of
$\T$ that contradicts Lemma~\ref{l.cycle-lda}. \qed


\subsection{Decomposing the input-$k$-ambiguous transducer $\Tpsi$}
\label{s.dka}

As observed in Section~\ref{s.N-sep}, $\T$ and $\Tpsi$ are equivalent
(for every $N$).
Thus, a decomposition of $\Tpsi$ is also a
decomposition of $\T$.

\begin{figure}
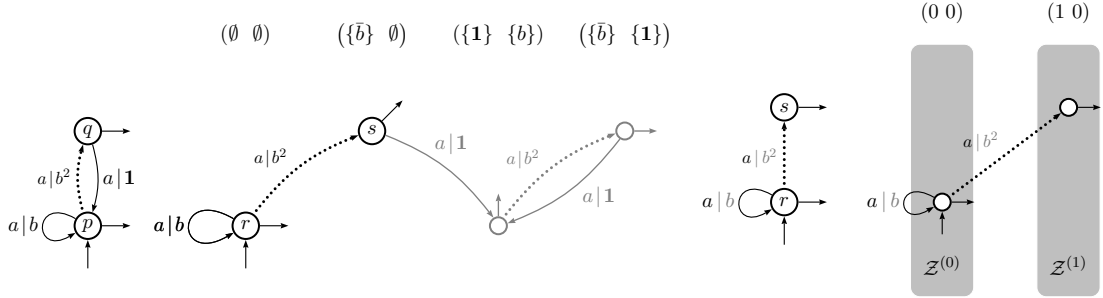

  \centering
  \TinyPicture
  \subfigure[A lag separation covering $\Tpsc$ with $N = 1$.
  The $\jspart{\Delta_1}^Q$-vectors (vertical projection) are indexed by
  $\{p, q\}$, in that order.
  The dotted transition is larger than the solid one.
  The input-$2$-ambiguous (and
  equivalent to $\T$) subtransducer $\Vc_1$ is reduced to the states
  $\{r, s\}$.]{
    \VCDraw{%
      \begin{VCPicture}{(-2,-2)(18,6)} 
        %
        %
        \VCPut[90]{(0,0)}{
          \State[p]{(0,0)}{p}
          \State[q]{(3,0)}{q}
          \Initial[w]{p}
          \Final[s]{p}
          \Final[s]{q}
        }
        \dotted{\ArcL[.5]{p}{q}{\IOL{a}{b^2}}}
        \ArcL[.5]{q}{p}{\IOL{a}{\ewfg}}
        \LoopW[.5]{p}{\IOL{a}{b}}
        %
        %
        \ChgStateLineStyle{none}
        \State[\left (\emptyset \ \ \emptyset \right)]{(5,6)}{VV}
        \State[\left (\{\bar{b}\} \ \ \emptyset \right)]{(9,6)}{AV}
        \State[\left (\{\ewfg\} \ \ \{b\} \right)]{(13,6)}{ea}
        \State[\left (\{\bar{b}\} \ \ \{\ewfg\} \right)]{(17,6)}{Ae}
        %
        %
        \ChgStateLineStyle{solid}
        \State[r]{(5,0)}{pVV2}
        \State[s]{(9,3)}{qAV2}
        \DimState
        \SmallState
        \State{(13,0)}{pea2}
        \State{(17,3)}{qAe2}
        \RstState
        \RstEdge
        \Initial[s]{pVV2}
        \Final[e]{pVV2}
        \Final[e]{qVV2}
        \Final[ne]{qAV2}
        \Final[e]{pVa2}
        \DimEdge
        \Final[n]{pea2}
        \Final[e]{qAe2}
        \RstEdge
        \dotted{\ArcL{pVV2}{qAV2}{\IOL{a}{b^2}}}
        \LoopW[.5]{pVV2}{\IOL{a}{b}}
        \LoopW[.5]{pVV2}{\IOL{a}{b}}
        \DimEdge
        \ArcL{qAV2}{pea2}{\IOL{a}{\ewfg}}
        \ArcL{qAe2}{pea2}{\IOL{a}{\ewfg}}
        \dotted{\ArcL{pea2}{qAe2}{\IOL{a}{b^2}}}
      \end{VCPicture}
    }
    \label{f.dec-lsc}
  }
  \subfigure[The lifted transducers $\Asil^{(0)}$ and $\Asil^{(1)}$ from 
  a $2$-skimming of the input automaton of $\Vc_1$.
  The choices of final states yield 
  $\CompAuto{\Asil^{(0)}} \colon a^n \mapsto b^n \, (n \geq 0)$ and
  $\CompAuto{\Asil^{(1)}} \colon a^n \mapsto b^{n + 1} \, (n > 0)$.]{
    \VCDraw{%
      \begin{VCPicture}{(-3,-3)(11,5.5)}
        %
        %
        \rput{0}(4,3){\psframe[linewidth=0pt,linecolor=white,fillstyle=solid,fillcolor=lightgray,framearc=.3](0,2)(2,-6)}
        \rput{0}(8,3){\psframe[linewidth=0pt,linecolor=white,fillstyle=solid,fillcolor=lightgray,framearc=.3](0,2)(2,-6)}
        \ChgStateLineStyle{none}
        \ChgStateFillColor{lightgray}
        \State[\Asil^{(0)}]{(5,-2)}{B0}
        \State[\Asil^{(1)}]{(9,-2)}{B1}
        \ChgStateLineStyle{solid}
        \ChgStateFillColor{white}
        %
        %
        \RstStateLineStyle
        \State[r]{(0,0)}{P}
        \State[s]{(0,3)}{Q}
        \RstEdgeLineStyle
        \Initial[s]{P}
        \Final[e]{P}
        \Final[e]{Q}
        \LoopW[.5]{P}{a \, \color{gray}{| \, b}}
        \dotted{\EdgeL{P}{Q}{a \, \color{gray}{| \, b^2}}}
        %
        %
        \ChgStateLineStyle{none}
        \State[(0 \,\, 0)]{(5,6)}{zz}
        \State[(1 \,\, 0)]{(9,6)}{uz}
        %
        %
        \ChgStateLineStyle{solid}
        \SmallState
        \State{(5,0)}{Pzz}
        \State{(9,3)}{Quz}
        \RstEdgeLineStyle
        \Initial[s]{Pzz}
        \Final[e]{Pzz}
        \Final[e]{Quz}
        \LoopW[.5]{Pzz}{a \, \color{gray}{| \, b}}
        \dotted{\EdgeL{Pzz}{Quz}{a \, \color{gray}{| \, b^2}}}
      \end{VCPicture}
    }
    \label{f.dec-ms}
  }
  \vspace*{-0.3cm}

  \caption{A decomposition of the $2$-valued transducer $\Tc$ of
    Figure~\ref{f.T}.}
  \label{f.dec}
\end{figure}

Take $N = Ln^{k   + 1}$ and let $\Aka$ be the underlying input
automaton of $\Tpsi$.
It is straightforward to decompose $\Tpsi$ by applying the
multi-skimming covering on $\Ac$.
By Proposition~\ref{p.N-sep-k-amb}, $\Aka$ is $k$-ambiguous,
hence the multi-skimming covering yields unambiguous automata
$\Bc_k^{(0)}, \dots, \Bc_k^{(k - 1)}$ which are immersions in $\Aka$,
and whose successful computations are in bijection with those of
$\Aka$.
By lifting to the transitions of $\Bc_k^{(0)}, \dots, \Bc_k^{(k - 1)}$
the corresponding outputs in $\Tpsi$ of the projected ones, we obtain
unambiguous transducers $\Asil^{(0)}, \dots, \Asil^{(k - 1)}$ whose
union is equivalent to $\Tc$.
Figure~\ref{f.dec} shows an example with a given ordering
for each covering.
Other decompositions are obtained by varying these orderings.

The number of states of the decomposition depends
on the following parameters of $\T$: $n$ (number
of states), $h$ (cardinality of the output alphabet), $L$
(maximal of the lengths of the outputs of transitions) and $k$
(valuedness).
We claim:

\begin{property}
  Each transducer $\Asil^{(i)}$ has at most $2^{\Oc(hLk^4n^{k + 4})}$
  useful states.
  \label{py.size-dec}
\end{property}

The proof is based on a fine analysis of the 
useful states of $\Asil^{(i)}$ and goes as follows.
Let $X$ be the set of useful states of $\Tpsi$ (as said
in Section~\ref{s.N-sep}, $X \subseteq Q \times
\jsPartK{\Delta_N}{k}^Q$).
Each transducer $\Asil^{(i)}$ is obtained by the multi-skimming
covering of the underlying input automaton of $\Tpsi$, hence
its states belong to $X \times \Nmbb_k^X$ (assuming that $\Asil^{(i)}$
was built on the trim part of $\Tpsi$).
By the stated properties of the constructions, we can derive that
if\footnote{Capital letters are used in order to
  distinguish the states of $\Asil^{(i)}$ from the states
  of other automata.}
$(P, \Nvec{V})$ is useful in $\Asil^{(i)}$, then  
$\Nvec{V}$ has at most $kn$ entries different from $0$.
In other words, the set of coordinates of $\Nvec{V}$ having a nonzero
value belongs to $\jsPartK{X}{kn}$.
There are $k$ possible nonzero values for each such coordinate, namely $\{1,
\dots, k - 1, \omega\}$, thus the number of
useful states of $\Asil_i$ is at most $\card{X} \cdot
\card{\jsPartK{X}{kn}} \cdot k^{kn}$.
To conclude, it remains to use the discussion on the number of useful
states of $\Tpsi$ at the end of Section~\ref{s.N-sep}:
we have that $\card{\jsPartK{X}{kn}} \leq
\card{X}^{(kn)^2}$, and by Property~\ref{py.size-dec-1}, $\card{X}
\leq 2^{2hNk^2n}$.
With $N = n^{k + 1}L$, we obtain the bound of $2^{\Oc(hLk^4n^{k +
    4})}$ states.

\subsection{The morphic decomposition theorem}
\label{s.gen-dec}

We turn now to Theorem~\ref{t.gen-weber}, the proof of which goes in
four steps.
First, we construct a $k$-valued transducer $\T$
realising the composition $\tau \theta$.
This is done by relabelling the transitions of the transducer $\Sc$
realising $\tau$: every transition $p \comput{a | x} q$ of $\Sc$
is replaced by $p \comput{a | x\theta} q$.
Next, $\T$ is decomposed into $k$ unambiguous transducers
$\Asil^{(0)}, \dots, \Asil^{(k - 1)}$.
These transducers are immersions in $\Tpsi$ and, by composition of
morphisms, also in $\T$; but it may be the case that not every
successful computation of $\Tc$ is projected by some successful
one in the union of the $\Asil^{(i)}$.
The third and crucial step (described more precisely below)
consists, roughly speaking, to \emph{stick}
the successful computations of $\T$ to the transducers
$\Asil^{(0)}, \dots, \Asil^{(k - 1)}$ in order to obtain
equivalent (thus functional) transducers $\Wc^{(0)},
\dots, \Wc^{(k - 1)}$, not necessarily unambiguous, whose
successful computations project on the whole set of successful
computations of $\T$.
Finally, the transitions of each $\Wc^{(i)}$ are relabelled
in order to construct an immersion of $\Sc$: $e: p
\comput{a | y} q$ in $\Wc^{(i)}$ projects on a transition $f$ of
$\T$; the label of $f$ is, by construction, of form $a | x\theta$; the
output $y$ of $e$ is replaced by $x$.
This yields $k$ transducers decomposing $\Sc$, not necessarily
functional, but whose compositions with $\theta$ are functional.

The definition of the transducers $\Wc^{(0)},
\dots, \Wc^{(k - 1)}$ is based on a
generalisation of the property of functional transducers that
the lag between every pair of successful computations with same
label is bounded by some integer (this appears implicitly in a
proof of~\cite{BCPS03}).
\begin{property}
  Let $N = n^{k + 1}L$ and $K = 2(k + 1)N$.
  If $\T$ is $k$-valued, then
  for every successful computation $c$ of $\T$ there exists a
  successful computation $D$ in $\Asil^{(0)} \cup \dots \cup \Asil^{(k
    - 1)}$ with same input, same output and such that
  $\lag{c}{D\varphi} <  K$. \qed
  \label{py.gen-weber}
\end{property}

We can obtain each $\Wc^{(i)}$ from the
product of $\T \times \Asil^{(i)}$ by the Lead or Delay action,
see~\cite{BCPS03} for details.
The part of this product restricted to states having Lead or Delay
in $\lagset{K}$ projects on the successful computations of
$\T$ with lag smaller than $K$ with some successful computation in
$\Asil^{(i)}$.
The number of states of $\Wc^{(i)}$ is bounded by $n \times M \times
\card{\lagset{K}}$, where $M$ is the number of
states of $\Asil{(i)}$.
This is again of order $2^{\Oc(hLk^4n^{k + 4})}$.

\end{document}